\documentclass[a4paper,fleqn,usenatbib]{mnras}

\usepackage{graphicx}	
\usepackage{amsmath}	
\usepackage{amssymb}	

\title[Representative AMRs]{Representative galaxy age-metallicity relationships}

\author[Piatti, Aparicio \& Hidalgo]{
Andr\'es E. Piatti$^{1,2}$\thanks{E-mail: andres@oac.unc.edu.ar (AEP)}, 
Antonio Aparicio$^{3,4}$ and  Sebasti\'an L. Hidalgo$^{3,4}$
\\
$^{1}$Consejo Nacional de Investigaciones Cient\'{\i}ficas y T\'ecnicas, Av. Rivadavia 1917, 
C1033AAJ, Buenos Aires, Argentina\\
$^{2}$Observatorio Astron\'omico, Universidad Nacional de C\'ordoba, Laprida 854, 5000, 
C\'ordoba, Argentina\\
$^{3}$Instituto de Astrof\'{\i}sica de Canarias, Calle V\'{\i}a L\'actea s/n, E-38200 La Laguna, Tenerife, Spain\\
$^{4}$Departamento de Astrof\'{\i}sica, Universidad de La Laguna, Avda. Astrof\'{\i}sico Fco. S\'anchez s/n, E-38206 La Laguna, Tenerife, Spain\\
}

\date{Accepted XXX. Received YYY; in original form ZZZ}

\pubyear{2016}

\begin{document}
\label{firstpage}
\pagerange{\pageref{firstpage}--\pageref{lastpage}}
\maketitle

\begin{abstract}
The ongoing surveys of galaxies and those for the next generation of telescopes will demand 
the execution of high-CPU consuming machine codes for recovering detailed
star formation histories (SFHs) and hence age-metallicity relationships (AMRs). We present 
here an expeditive  method which provides quick-look AMRs on the basis of representative ages and
metallicities obtained from colour-magnitude diagram (CMD) analyses. We have tested its 
perfomance by generating synthetic CMDs for a wide variety of galaxy SFHs.
The representative AMRs turn out to be reliable down to a magnitude limit with
a photometric completeness factor higher than $\sim$ 85 per cent, and trace the
chemical evolution history for any stellar population (represented by a mean age
and an intrinsic age spread) with a total mass within $\sim$ 40 per cent of the
more massive stellar population in the galaxy.
\end{abstract}




\section{Introduction}

The age-metallicity relationship (AMR) has fruitfully been employed to
study the chemical evolution of different galactic and extragalactic
stellar systems
\citep[e.g.][]{c82,catelanetal92,reyetal04}. When dealing with the composed
stellar population of a galaxy it has frequently been obtained from the
galaxy star formation history (SFH) 
\citep[e.g.][]{ah09,hidalgoetal11,deboeretal12,savinoetal15}.
If such SFHs are extracted from synthetic colour-magnitude
diagram-based tecniques \citep[e.g.][]{setal09,vandenbergetal15}, 
they usually demand the consumption of a lot of CPU-time.

In order to be more expeditive in obtaining AMRs from large photometric
databases, \cite{pietal12} developed a procedure based on the so-called 
{\it representative stellar populations}. The method has proved to be useful in comprehensively
tracing the AMRs of the Large and Small Magellanic Clouds (L/SMC) \citep{p12a,pg13}
and that of the Fornax spheroidal dwarf galaxy \citep{petal14c} as well. In this 
work, we explore the advantages and constraints of such method in 
obtained astrophysically meaningful AMRs in a more general framework, by applying 
it to synthetic photometric data sets generated for a wide variety of galaxy SFHs.

In Section 2 we describe the procedure of building galaxy AMRs from their representative
stellar populations and compare previous results from this method with those obtained
from independent ones. In Section 3 we present different synthetic SFHs used to probe the ability 
of the aforementioned technique, while in Section 4 we analyze and discuss its
usefulness in different galaxy formation and evolution scenarios.

\section{The representative stellar populations method}

Given the CMD of a composed stellar population, \cite{pietal12} assumed that the
observed main sequence (MS) in any galaxy field is a result of the superposition 
of MSs with different turnoffs (TOs) and constant luminosity functions (LFs). 
Hence, the difference between the number of stars
of two adjacent magnitude intervals gives the intrinsic number
of stars belonging to the faintest interval. Consequently, the
biggest difference is directly related to the most populated TO. \cite{getal03}
defined this TO as {\it representative} of a comparatively small field in the sky 
along the line of sight,
and soon after was used by \cite{petal03,petal03b,petal07d}, among others,
for revealing the primary trends of the stellar composition in different galaxy fields
in an efficient and robust way.
The definition of a representative TO could not converge
to any dominant TO (age) value if the stars in a given field came
from a constant star formation rate (SFR) integrated over all time. 
In such a case,
the difference between the number of stars of two adjacent
magnitude intervals would result in the same value for any magnitude interval.

The concept of representative TOs is directly applied to AMRs whenever 
the ages and metallicities used to build them come from representative values, i.e.,
ages and metallicities estimated by using the photometric information of
the prevailing stellar populations. These prevailing populations trace
the present-day AMR of a galaxy. They account for the most
important metallicity-enrichment processes that have undergone
in the galaxy lifetime. Minority stellar populations not following
these main chemical galactic processes are discarded. Therefore,
presently-subdominant populations in certain locations could have
been  present in the majority in the galaxy in the past, but were not 
considered. Note that any AMR built from representative ages and 
metallicities differs from those
derived from modelled SFHs in the fact that it does not include
complete information on all stellar populations, but accounts for
the dominant population present in each field. 

Red clumps (RCs) stars are usually used as standard candles for distance
determinations \citep{ps98,os02,s03}. However, they are also often used in age
estimates based on the magnitude difference $\delta$ between the
HB/clump and the MSTO for intermediate-age and old clusters \citep[see, e.g.][]{petal94}, 
since the RC mag is relatively invariant to population effects such as age and 
metallicity for such stars \citep{gs01}. Since the
MSTO magnitude is an excellent age indicator, so also is the difference (in magnitude)
MSTO $-$ RC. By assuming that the peak of the RC magnitude distribution of
a composed stellar population corresponds
to the most populous MSTO in the respective field, it is possible to estimate
representative
ages from composed stellar population CMDs, particularly for ages older than 1 Gyr.
For younger ages the magnitude of the He-burning stage varies
with age for such massive stars. Note that this age
measurement technique does not require absolute photometry
and is independent of reddening and distance as well. An
additional advantage is that it is not needed to go deep enough
to see the extended MS of the representative stellar population
but only slightly beyond its MSTO.

As for the representative metallicities, they can be derived following similar
precepts, i.e., by identifying the prevailing stellar population
(the more numerous one) in the CMD or Hess diagram of a particular field. For instance, 
if the position, slope and/or shape of the red giant branch (RGB) is used as 
a metallicity indicator \citep[e.g.][]{da90,gs99,valentietal04,chetal16}, the 
fiducial placement of the densest 
RGB path should be considered. Depending on the metallicity sensitivity of the
photometric system,  particular caveats should be taken into account
for those photometric metallicity estimation methods that show some slight
dependence with age \citep{getal03,os15}.

\begin{figure}
	\includegraphics[width=\columnwidth]{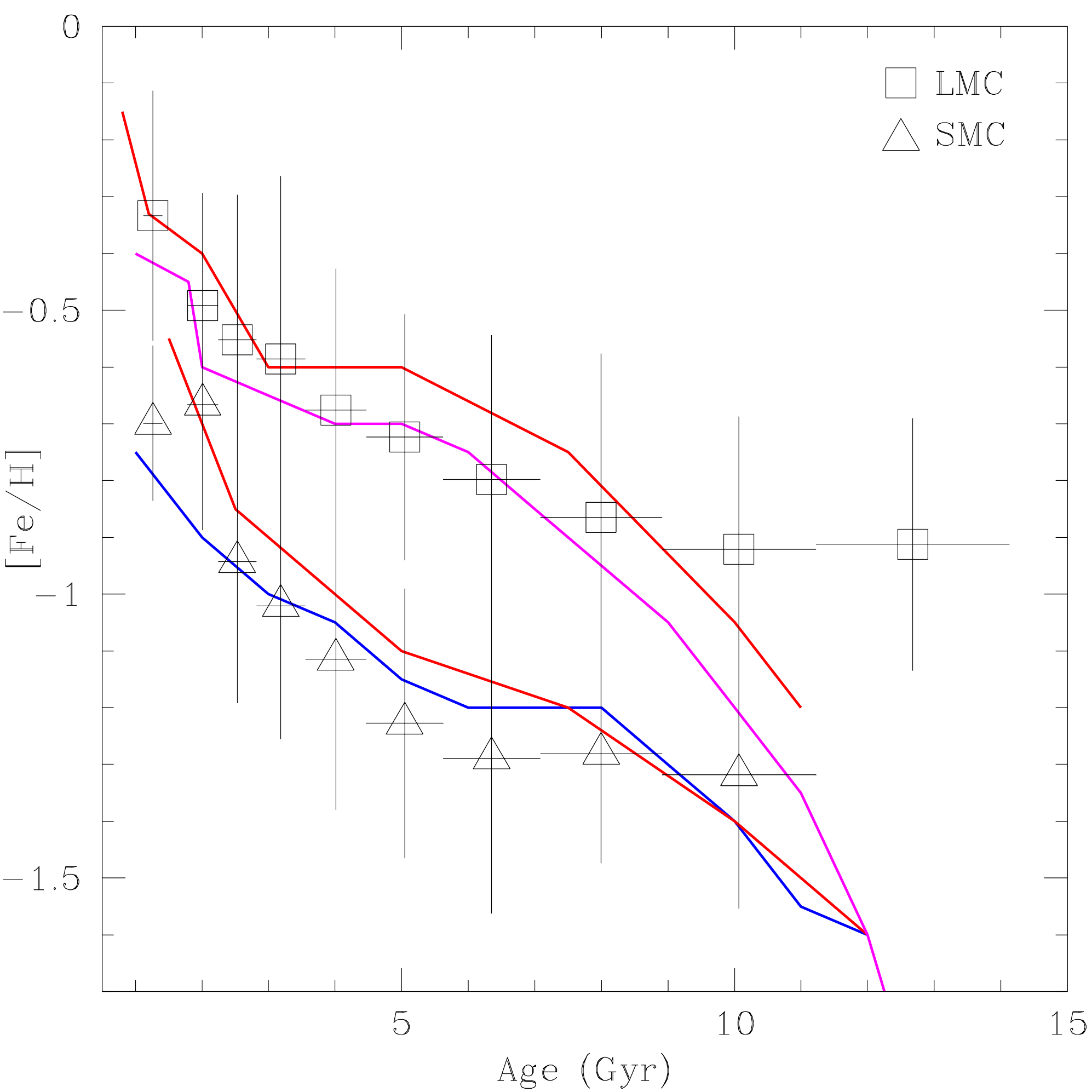}
    \caption{AMRs obtained by \citet[][ SMC, open triangles]{p12a} and 
\citet[][ LMC, open boxes]{pg13} from representative 
stellar populations. Error bars represent intrinsic dispersion of the representative 
 populations. The AMRs derived by \citet[][red lines]{retal12,retal15}, 
\citet[][magenta]{bt12} and \citet[][blue line]{cetal13b} 
are superimposed for comparison purposes.}
   \label{fig:fig1}
\end{figure}

\cite{p12a} and \cite{pg13} built representative AMRs for the SMC and LMC using 
some 3.3 and 5.5 million stars observed in the Washington $CT_1$ photometric system,
respectively, distributed throughout the main body of each galaxy. The representative 
MSTO magnitudes turned out to be on average $\sim$0.6 mag brighter than the mag for 
the 100\% completeness level of the respective fields,
so that they actually reached the TO of the representative population of each field.
Fig.~\ref{fig:fig1} reproduces 
their AMRs that trace the main features of the chemical enrichment experienced by 
these galaxies. Since them, some other independent AMRs have been obtained
by \citet[][$YJK_s$ survey]{retal12} and \citet[][theoretical model]{bt12} for the LMC, and
by \citet[][$HST$ photometry]{cetal13b} and \citet[][$YJK_s$ survey]{retal15} for the SMC, 
among others. Particularly, \citet{retal12,retal15} and  \citet{cetal13b} used
independent procedures to fit synthetic CMDs to the data. They have been overplotted
on Fig.~\ref{fig:fig1} for comparison purposes. As can be seen, all of them agree
reasonably well. 

\citet{petal14c} applied also the representative
stellar population method to study the AMR of the Fornax dwarf spheroidal galaxy
from $VI$ photometric data obtained with FORS1 at the VLT. According to \citet{delpinoetal13}, the derived representative $V$(MSTO) mags of each surveyed subregion are brighter than the $V$ mag at the 90\% completeness level, so that they actually
reached the MSTO of the representative oldest populations of the galaxy, particularly
 in those regions where the oldest populations are indeed the dominant population. 
When focusing on the most prominent stellar populations, they found that the derived AMRs 
are engraved by the evidence of an outside-in star formation process and suggested
for the first time a possible merger of two galaxies that would have
triggered a star formation bursting process that peaked between $\sim$6 and 9 Gyr ago, 
depending on the position of the field in the galaxy. Later on, other studies
showed similar outcomes \citep{hendricksetal14,delpinoetal15,wangetal16}, thus validating 
the representative stellar population method.

At this point a question arises unavoidably: since the concept {\it representative}
has associated a specific galaxy tracer (the representative stellar population), we
wonder whether there are scenarios in the galaxy formation and evolution for which
the {\it representative} tracer differs from considering as tracers the whole stellar populations. 
In other words, could the representive AMR be used to describe the
global chemical enrichment history of any galaxy? In the subsequent Section we 
thoroughly examine its scope and constraints.

\section{synthetic age-metallicity relationships}

\subsection{The models}

In order to test wether the representative AMRs are robust representations of the most significant age and metallicity present in CMDs we have carried out the following procedure. First, we have generated four synthetic stellar populations with different input SFRs and AMRs; second, we have simulated realistic observational effects in their CMDs, and finally, we have applied the representative AMR method to each of the synthetic CMDs. Comparison of the obtained results and the input data will provide the required information for the robustness test. A short description of the first and second steps follows.

\begin{figure*}
\includegraphics[width=\columnwidth]{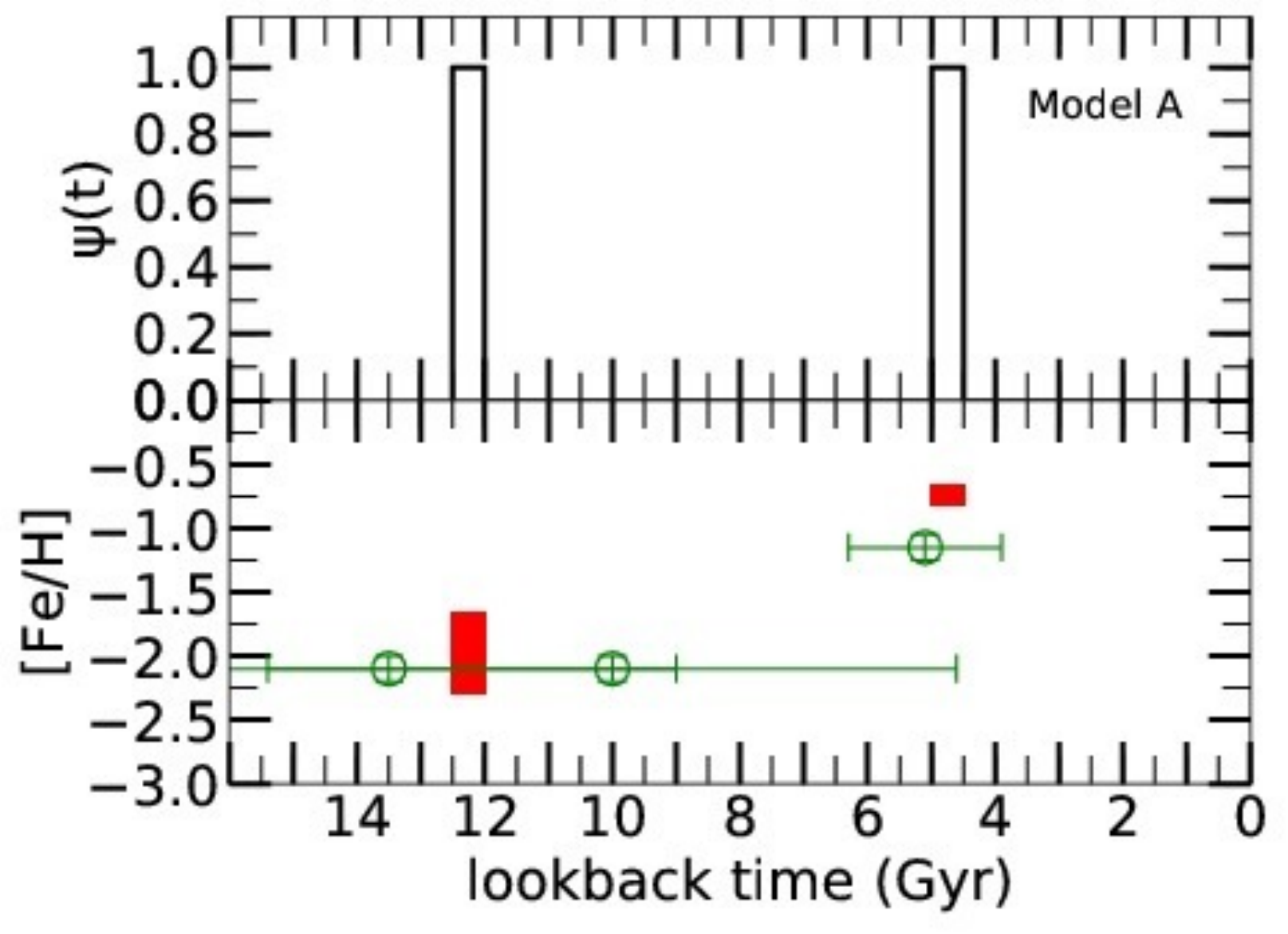}\\
\includegraphics[width=\columnwidth]{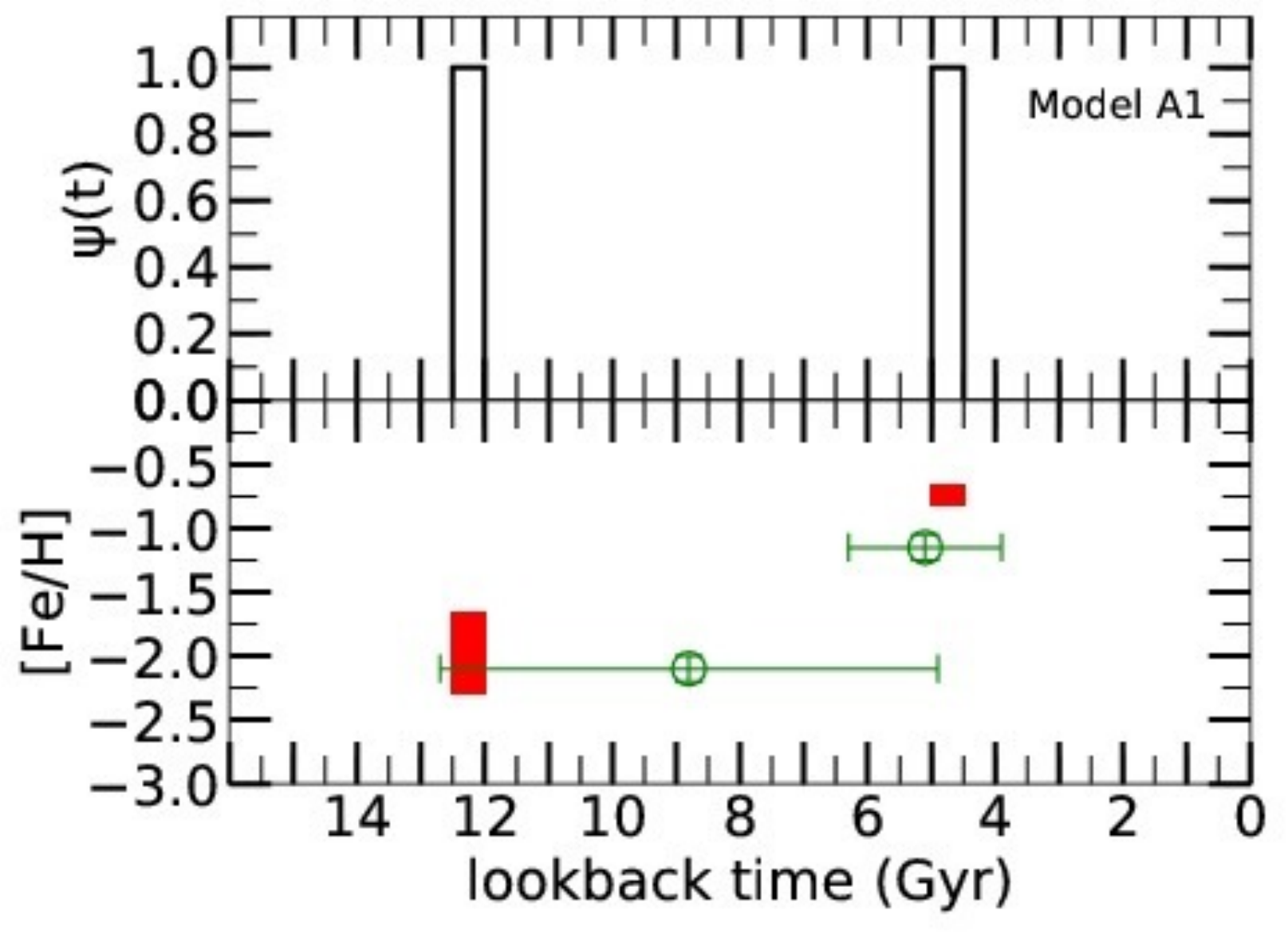}
\includegraphics[width=\columnwidth]{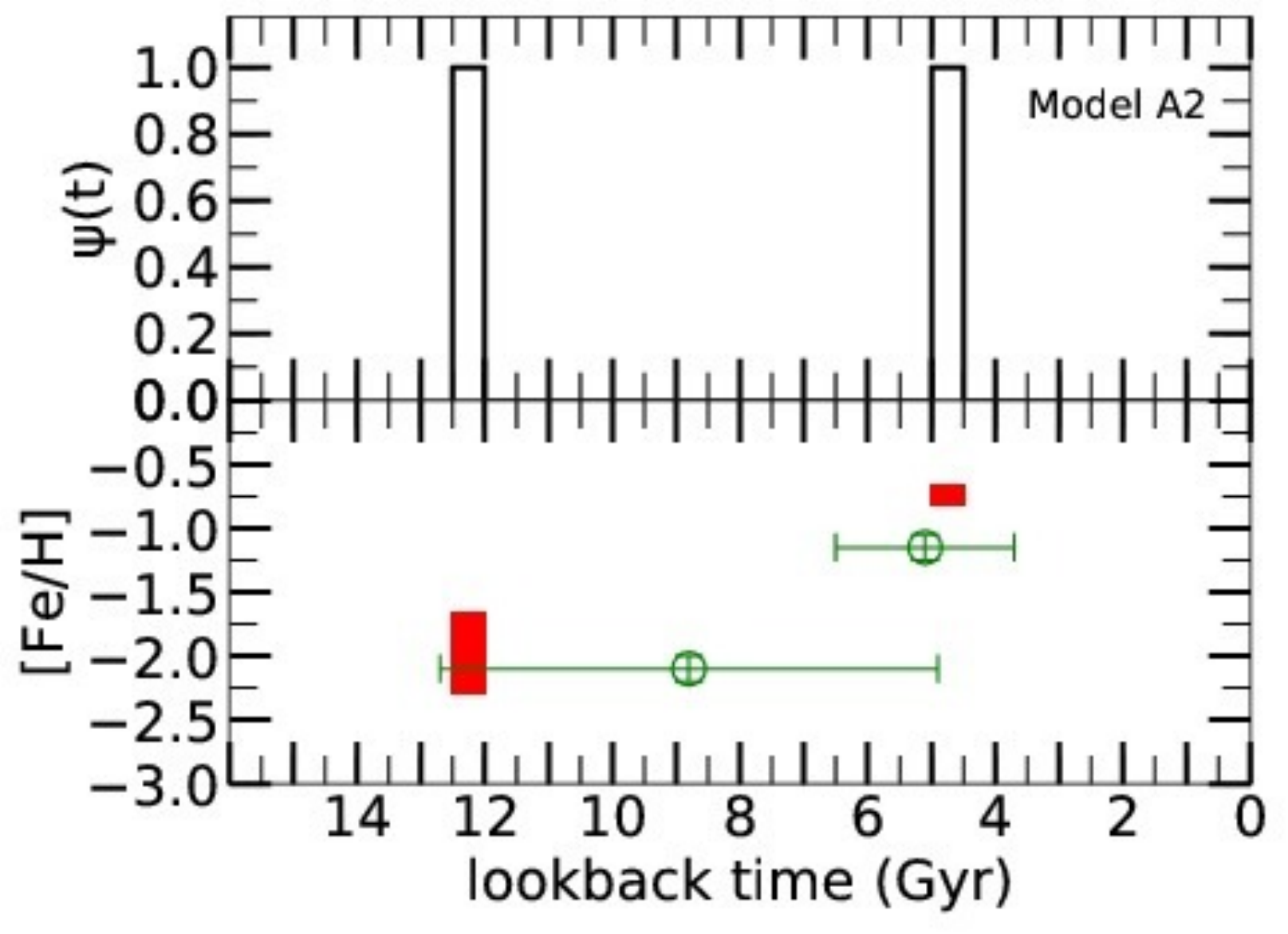}
\includegraphics[width=\columnwidth]{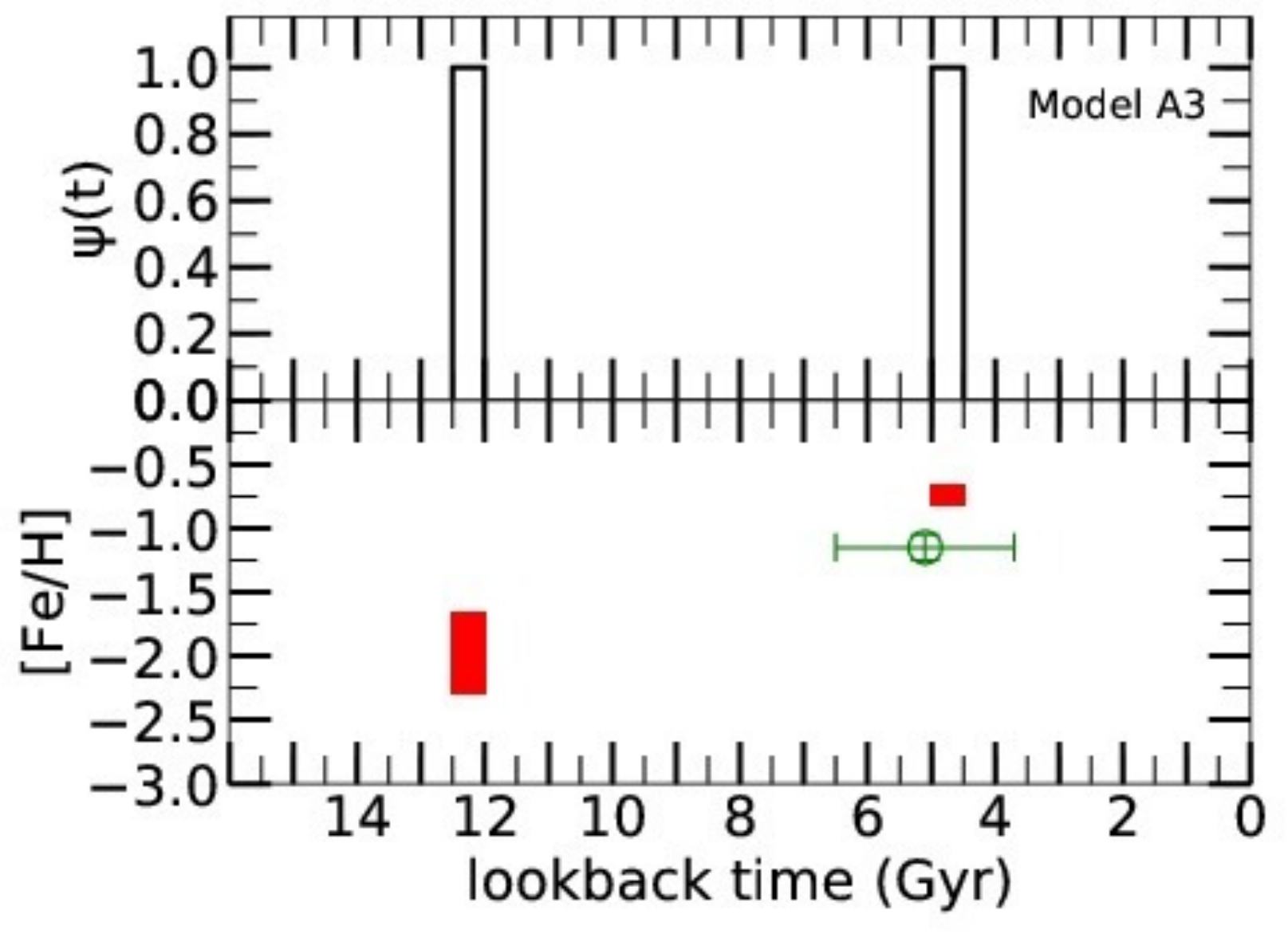}
\includegraphics[width=\columnwidth]{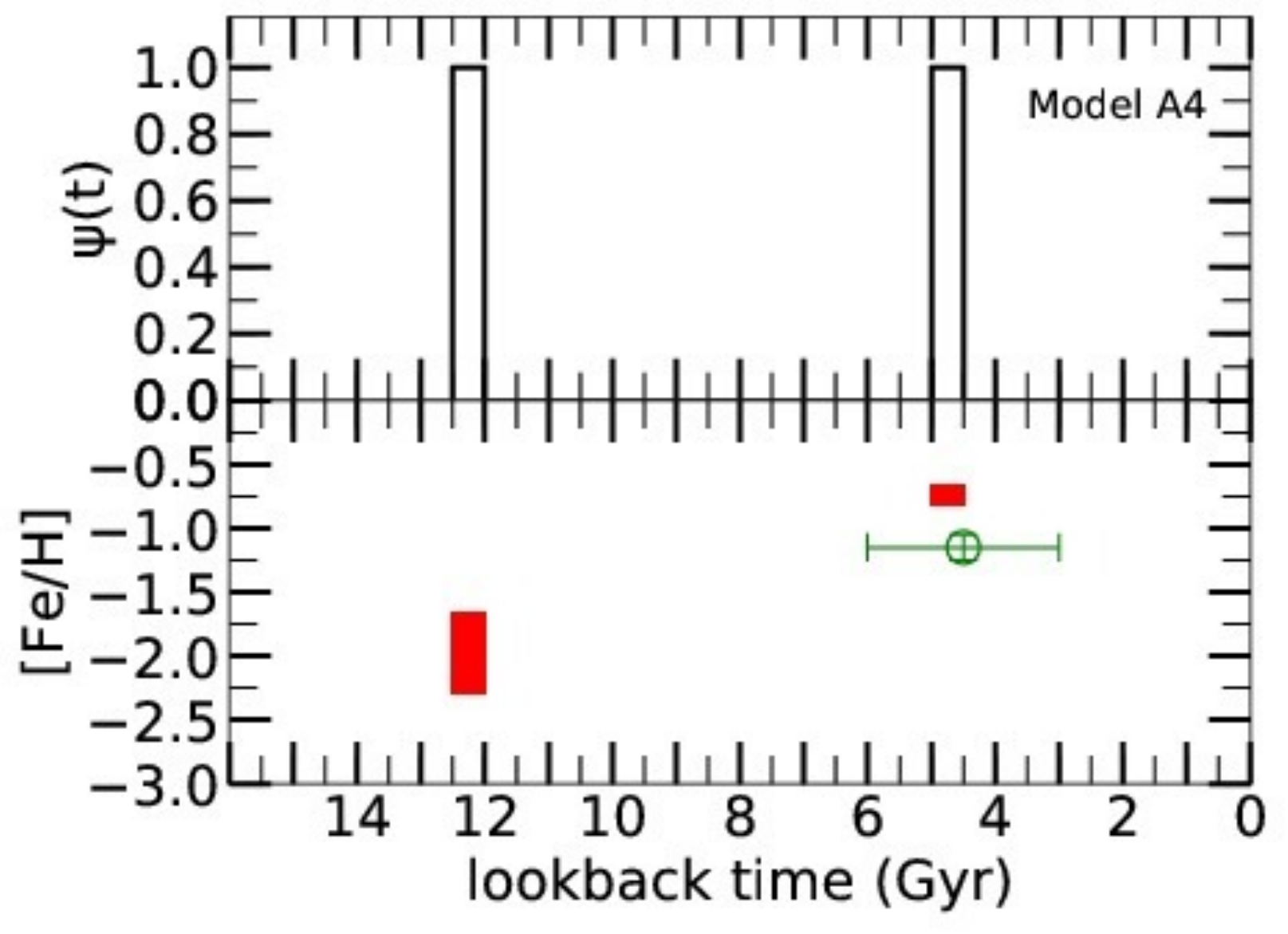}
    \caption{SFH and AMR generated for stellar population model  A.
 To transform $Z$ values to [Fe/H] ones we used the following relationship: [Fe/H] = log(Z/0.019) 
\citep{metal08}. The representative AMRs (Table~\ref{tab:table2}) are depicted with open circles for the input model free of observational effects and for the four scenarios obtained varying errors and completeness (see main text). In all the cases, the upper half panels represent the input SFH of the population model. The bottom half panels show, in red, the input AMR and, in green, the corresponding solutions for the representative AMR.}
   \label{fig:fig2}
\end{figure*}

\begin{figure*}
\includegraphics[width=\columnwidth]{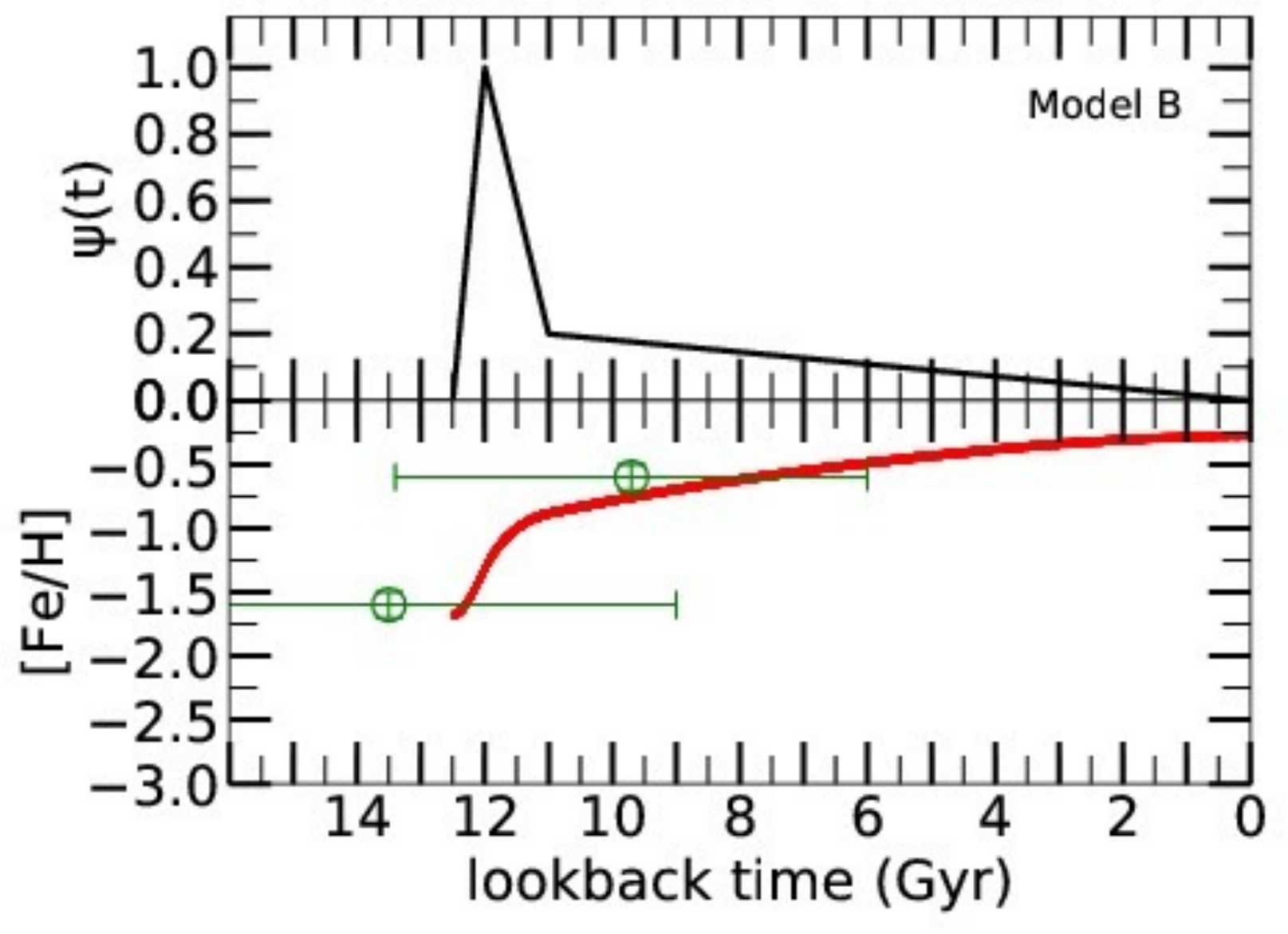}\\
\includegraphics[width=\columnwidth]{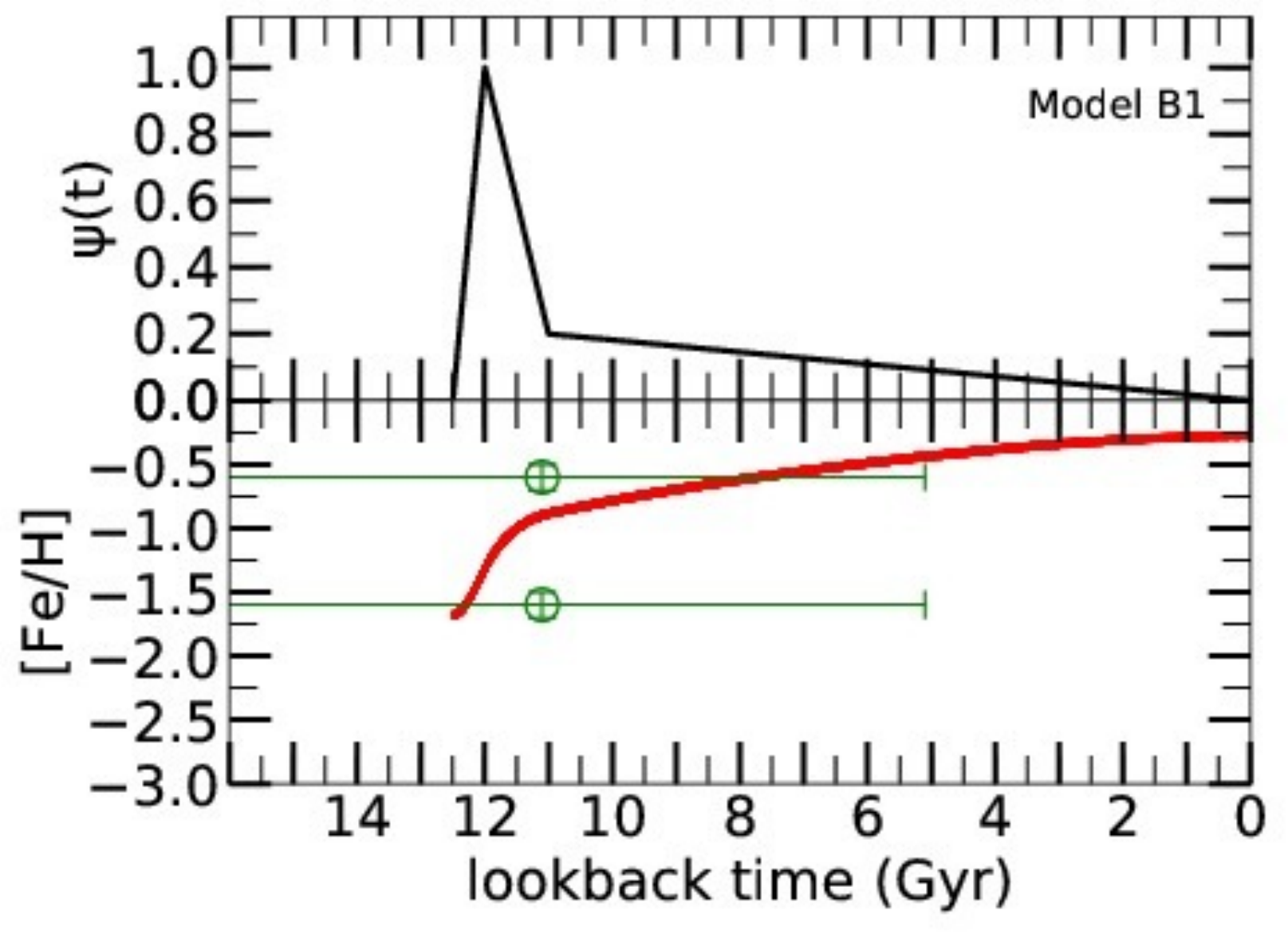}
\includegraphics[width=\columnwidth]{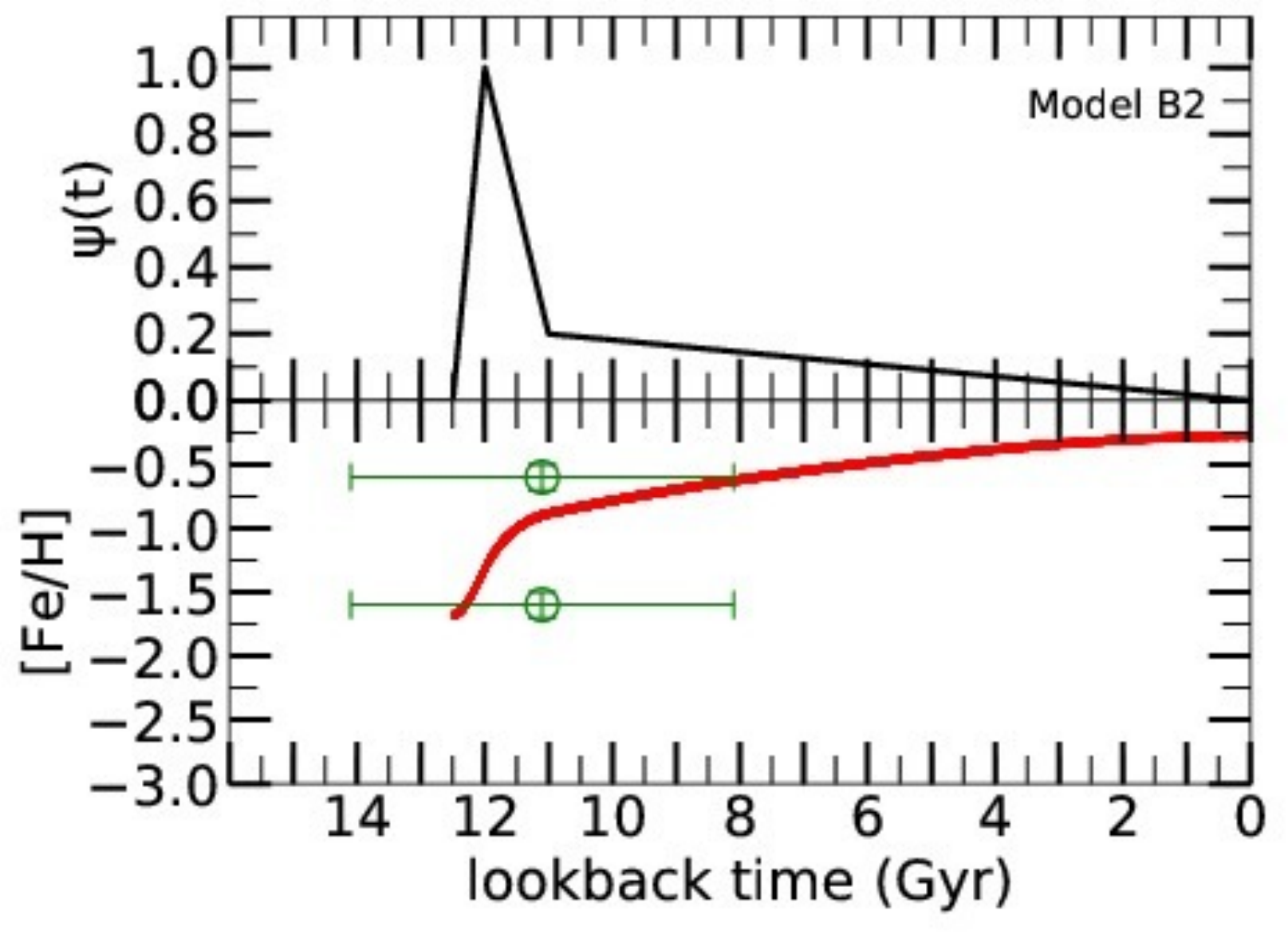}
\includegraphics[width=\columnwidth]{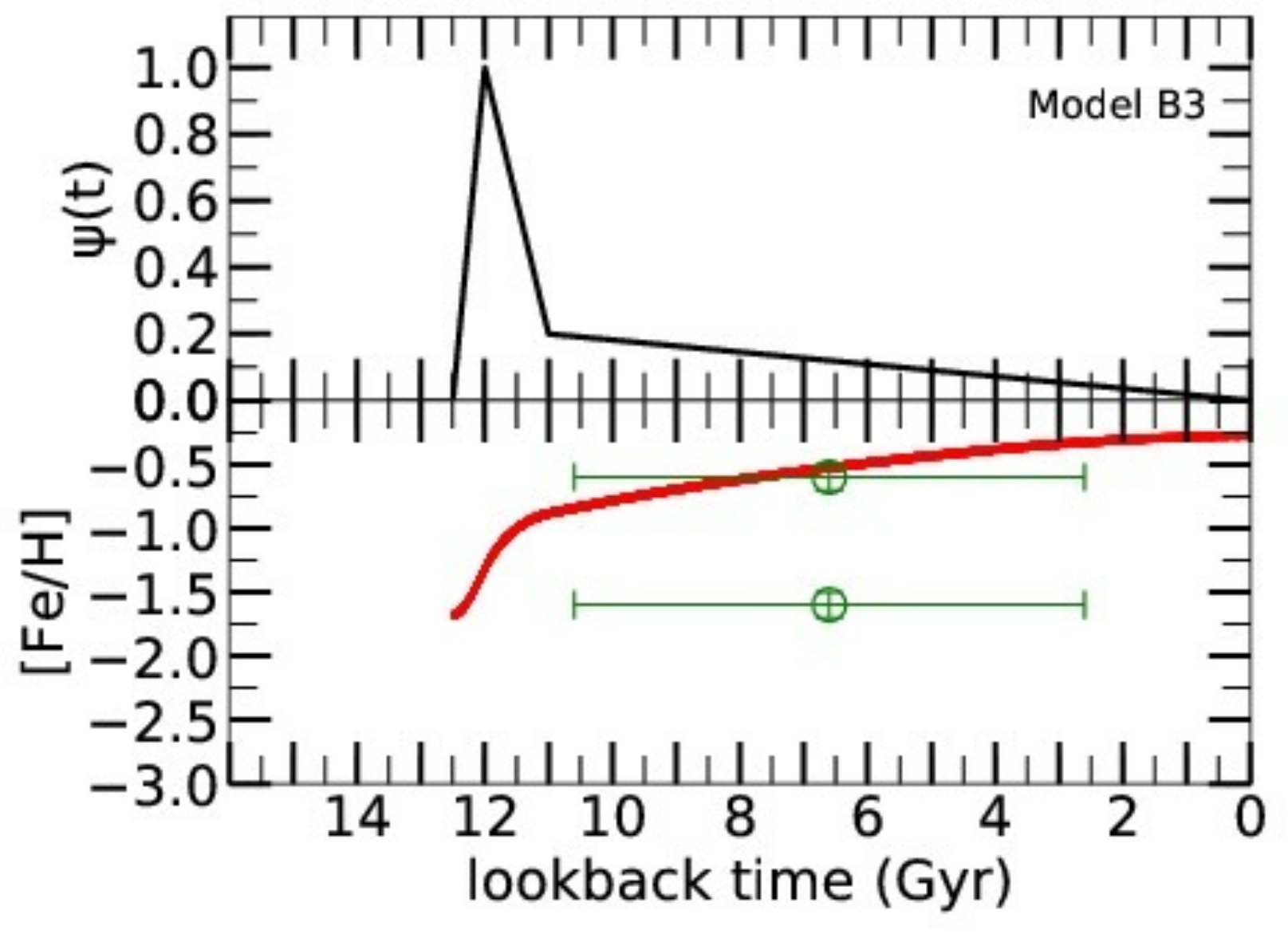}
\includegraphics[width=\columnwidth]{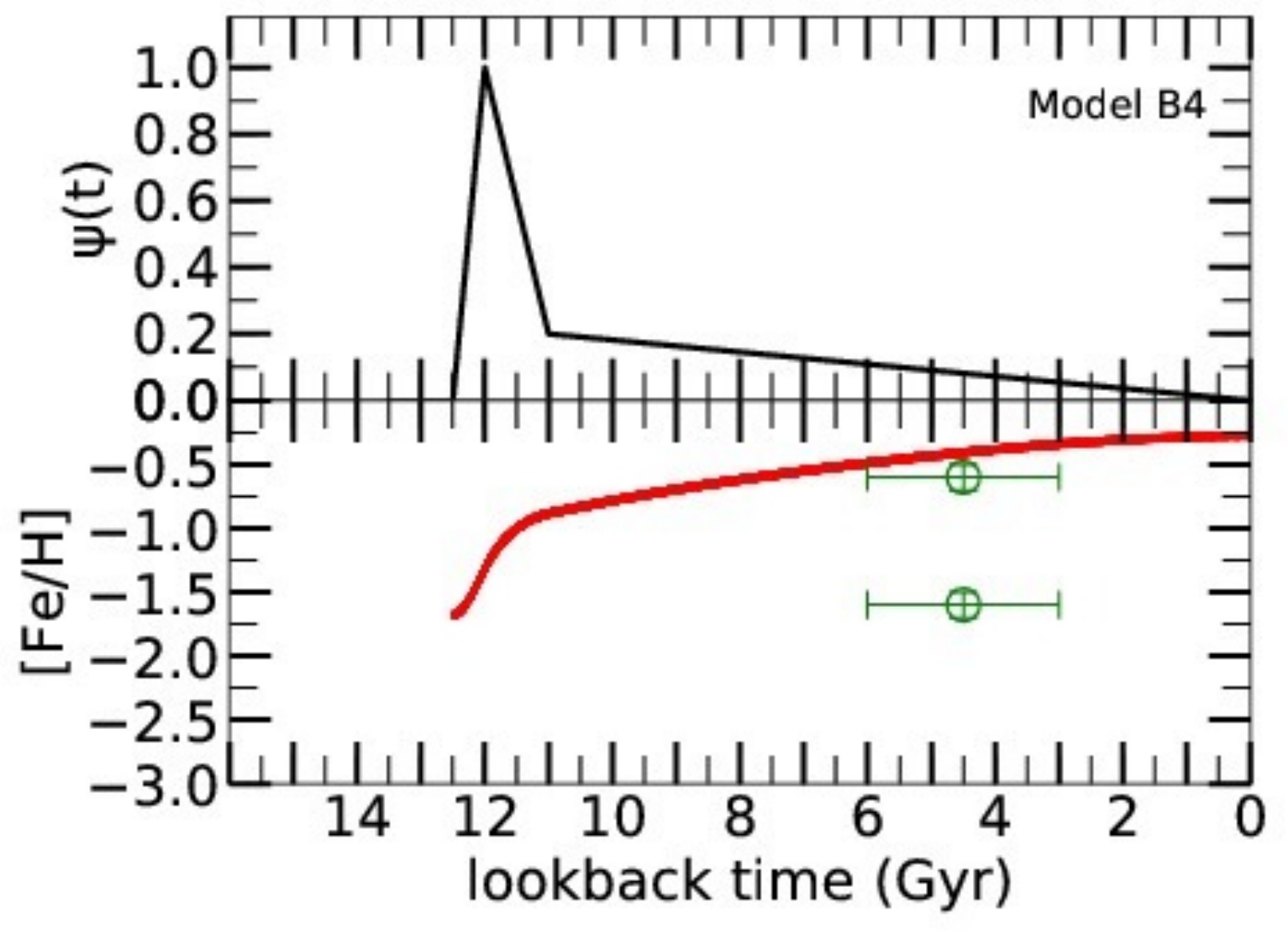}
    \caption{Same as Fig.~\ref{fig:fig2}, but for stellar population model  B.}
   \label{fig:fig3}
\end{figure*}

\begin{figure*}
\includegraphics[width=\columnwidth]{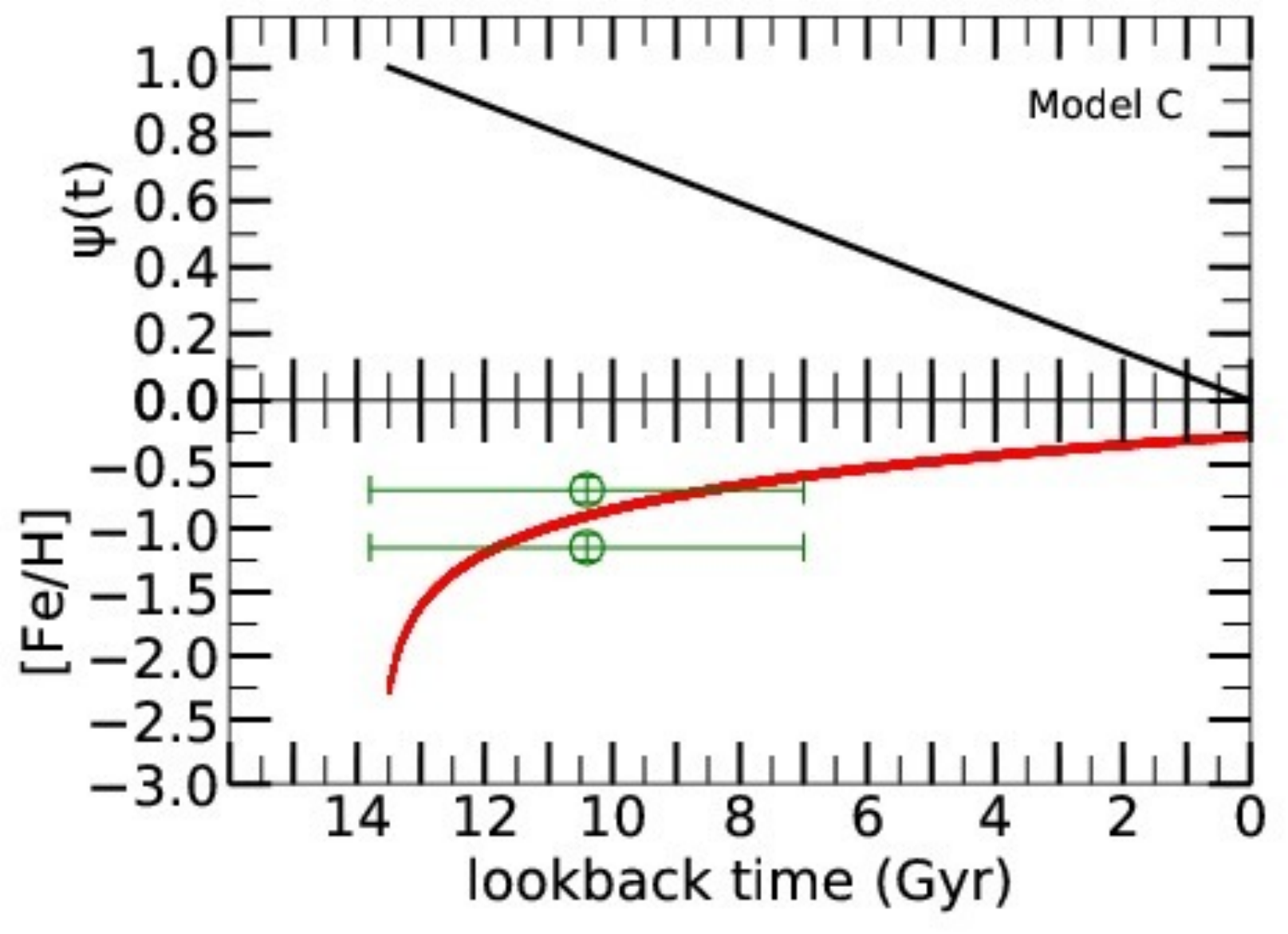}\\
\includegraphics[width=\columnwidth]{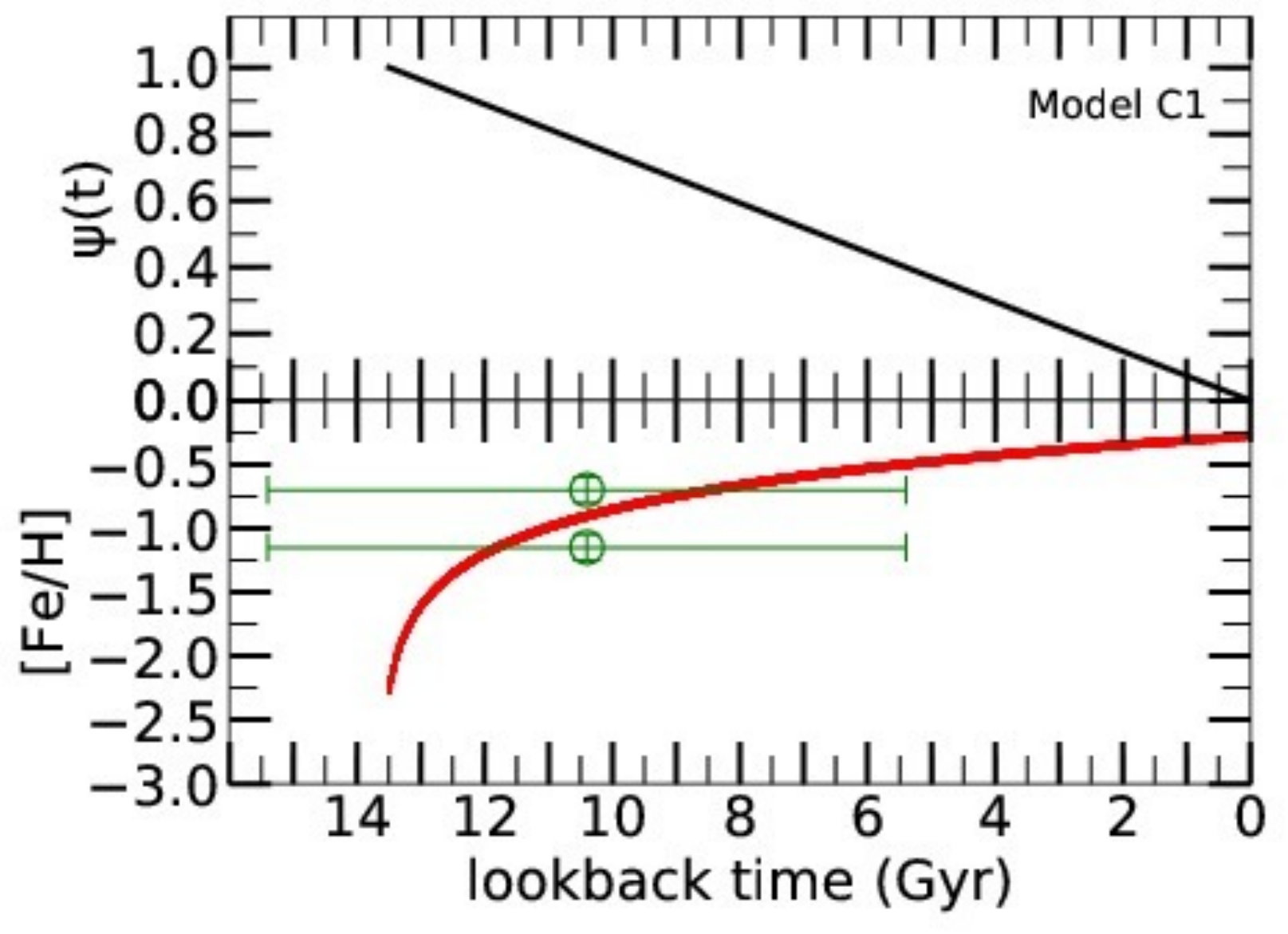}
\includegraphics[width=\columnwidth]{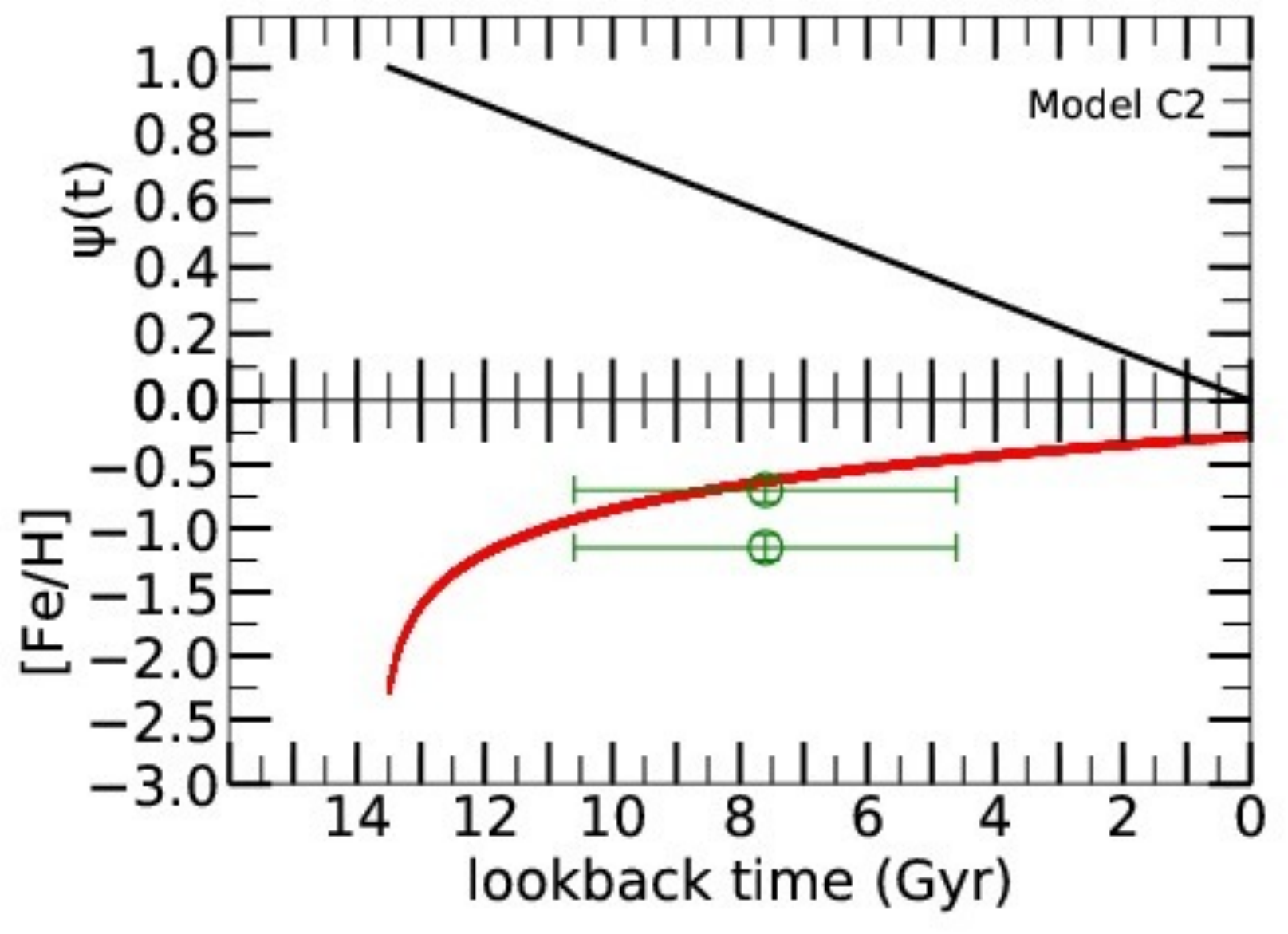}
\includegraphics[width=\columnwidth]{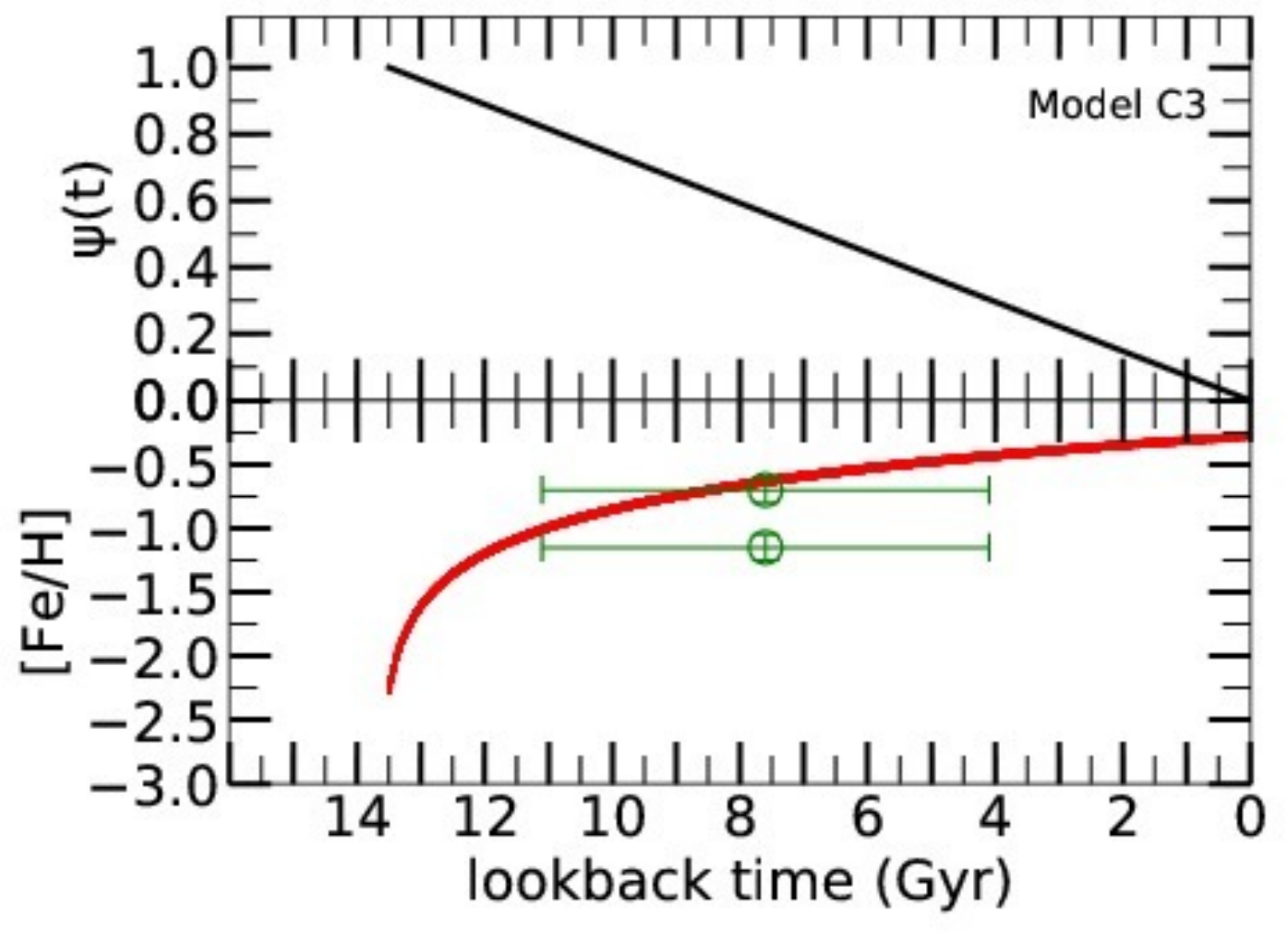}
\includegraphics[width=\columnwidth]{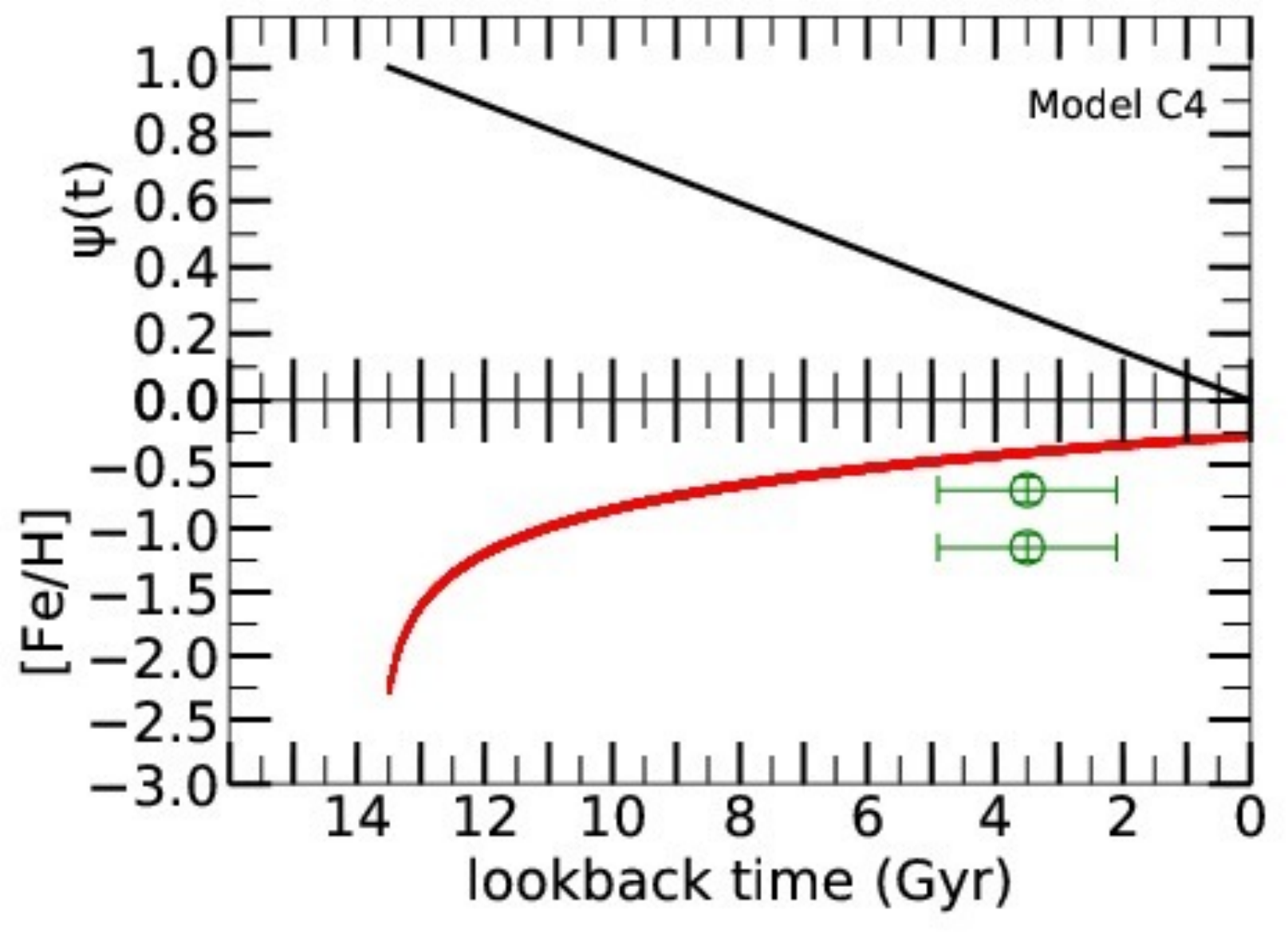}
    \caption{Same as Fig.~\ref{fig:fig2}, but for stellar population model  C.}
   \label{fig:fig4}
\end{figure*}

\begin{figure*}
\includegraphics[width=\columnwidth]{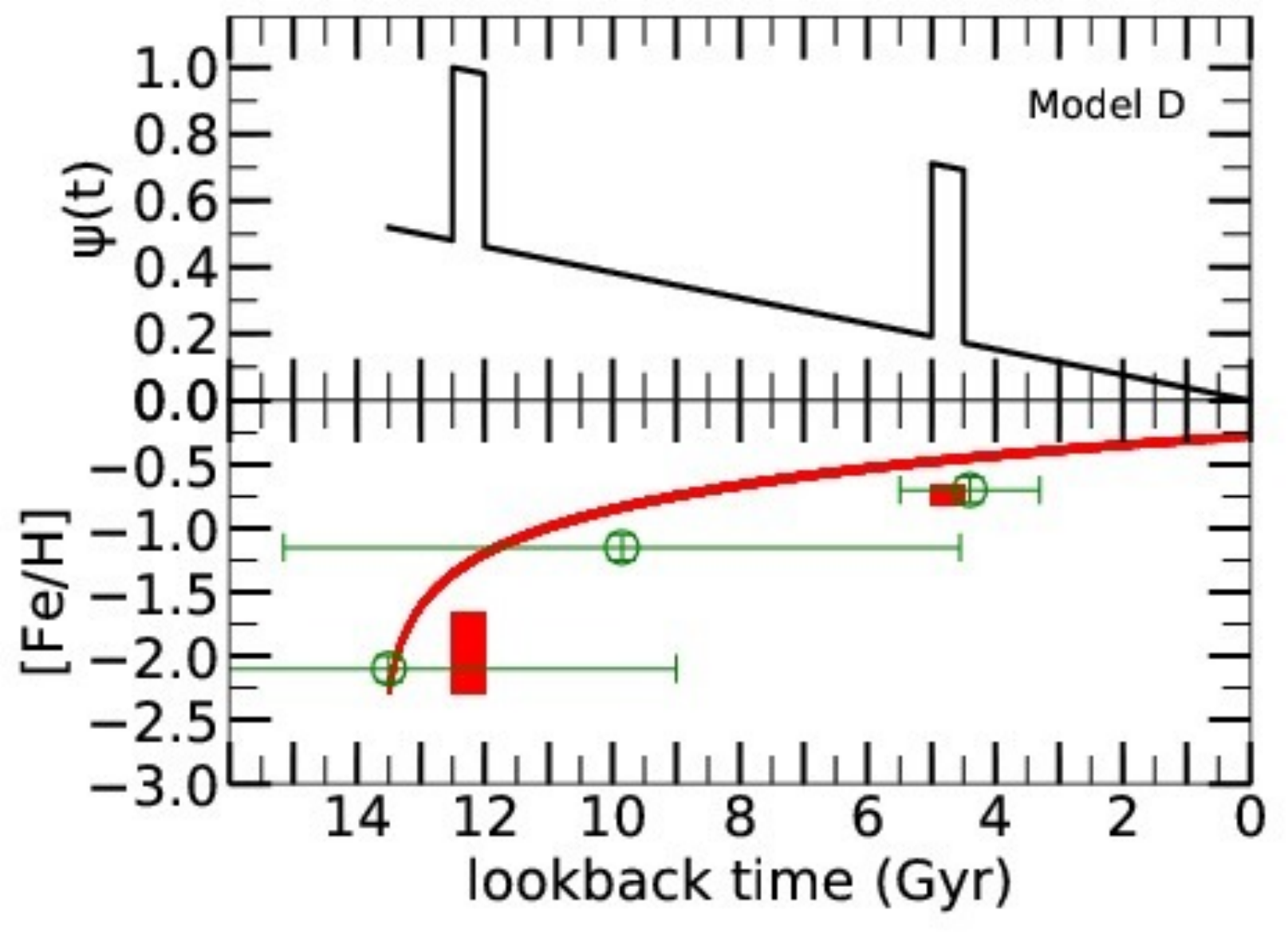}\\
\includegraphics[width=\columnwidth]{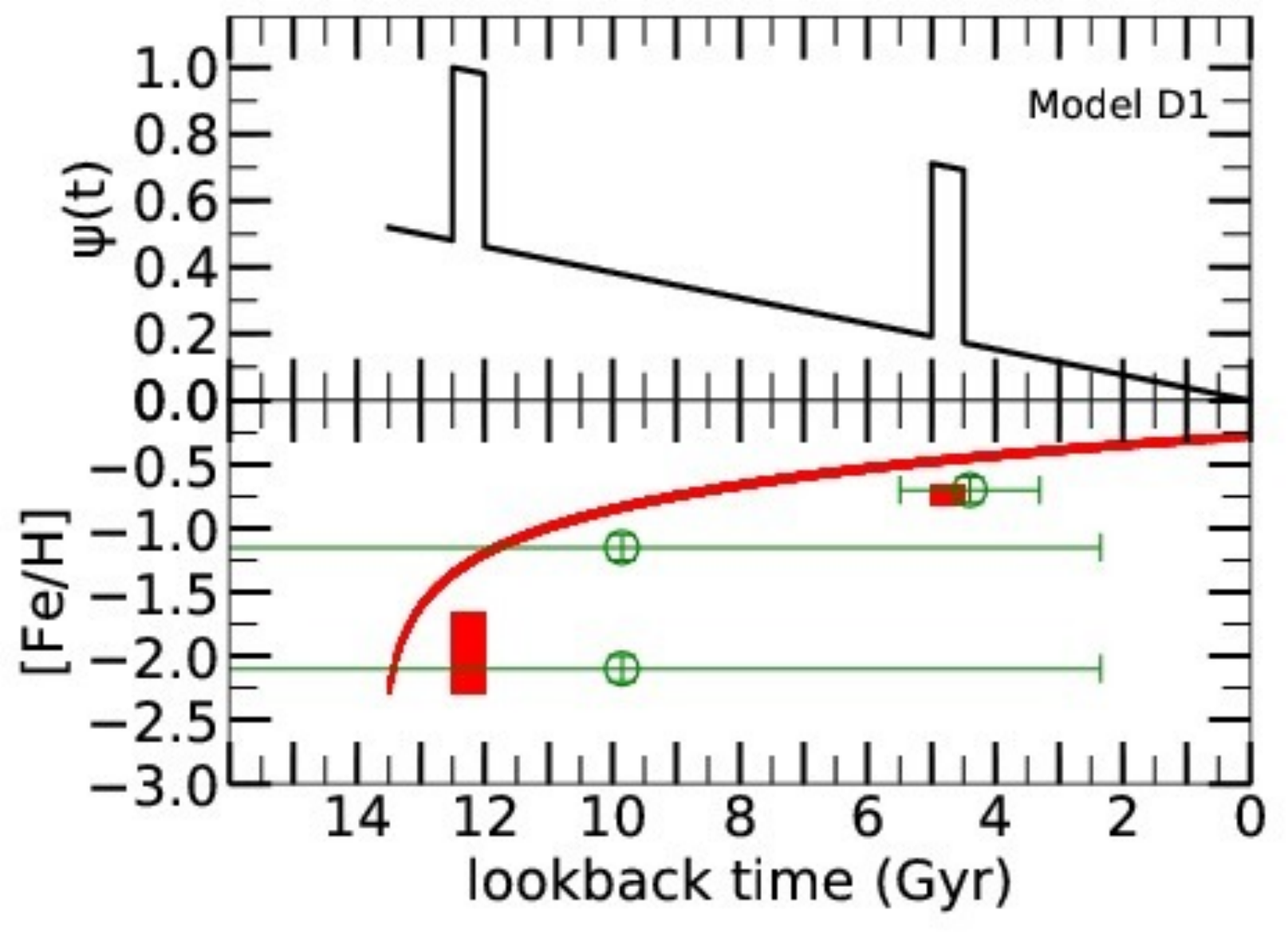}
\includegraphics[width=\columnwidth]{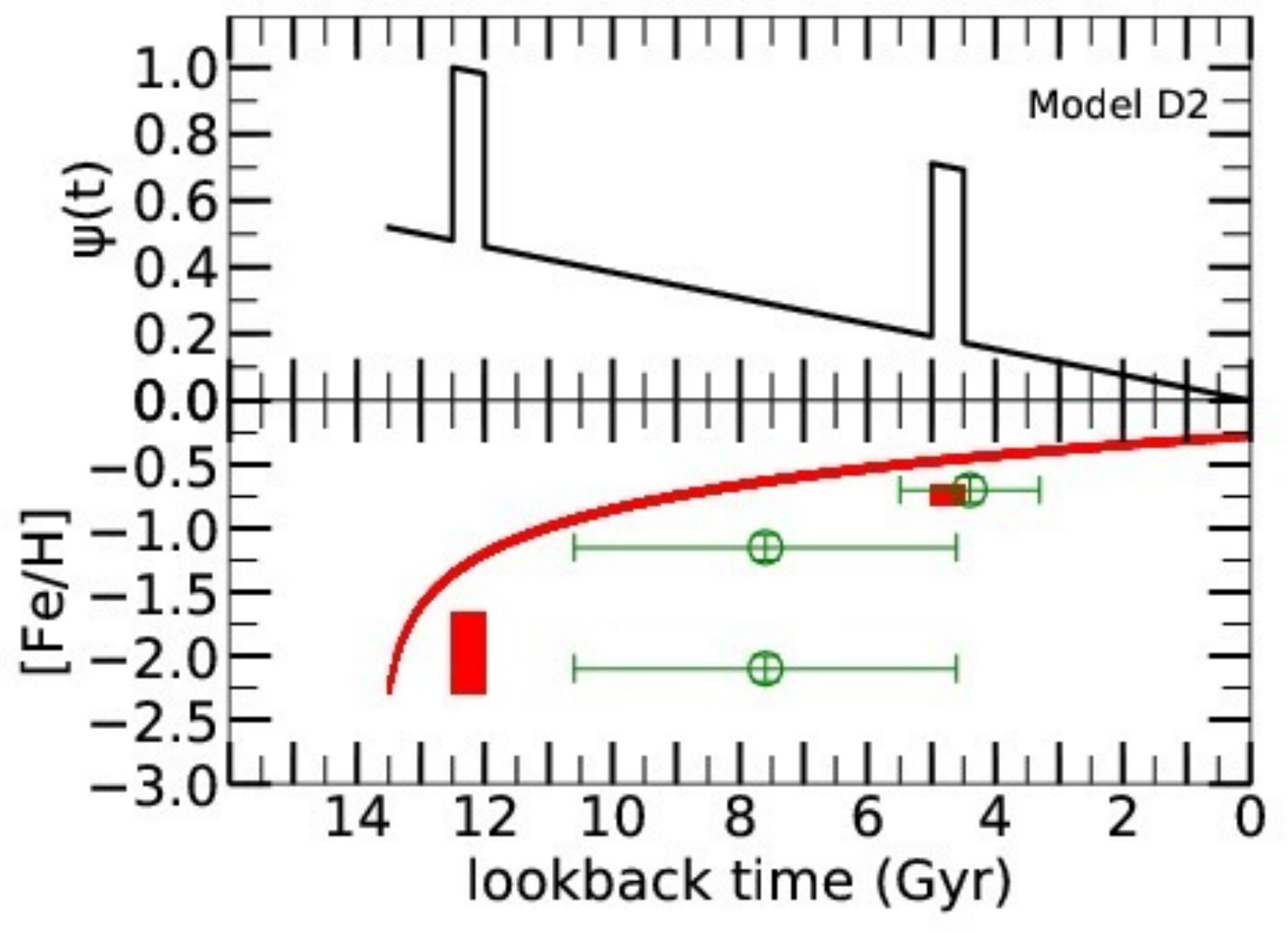}
\includegraphics[width=\columnwidth]{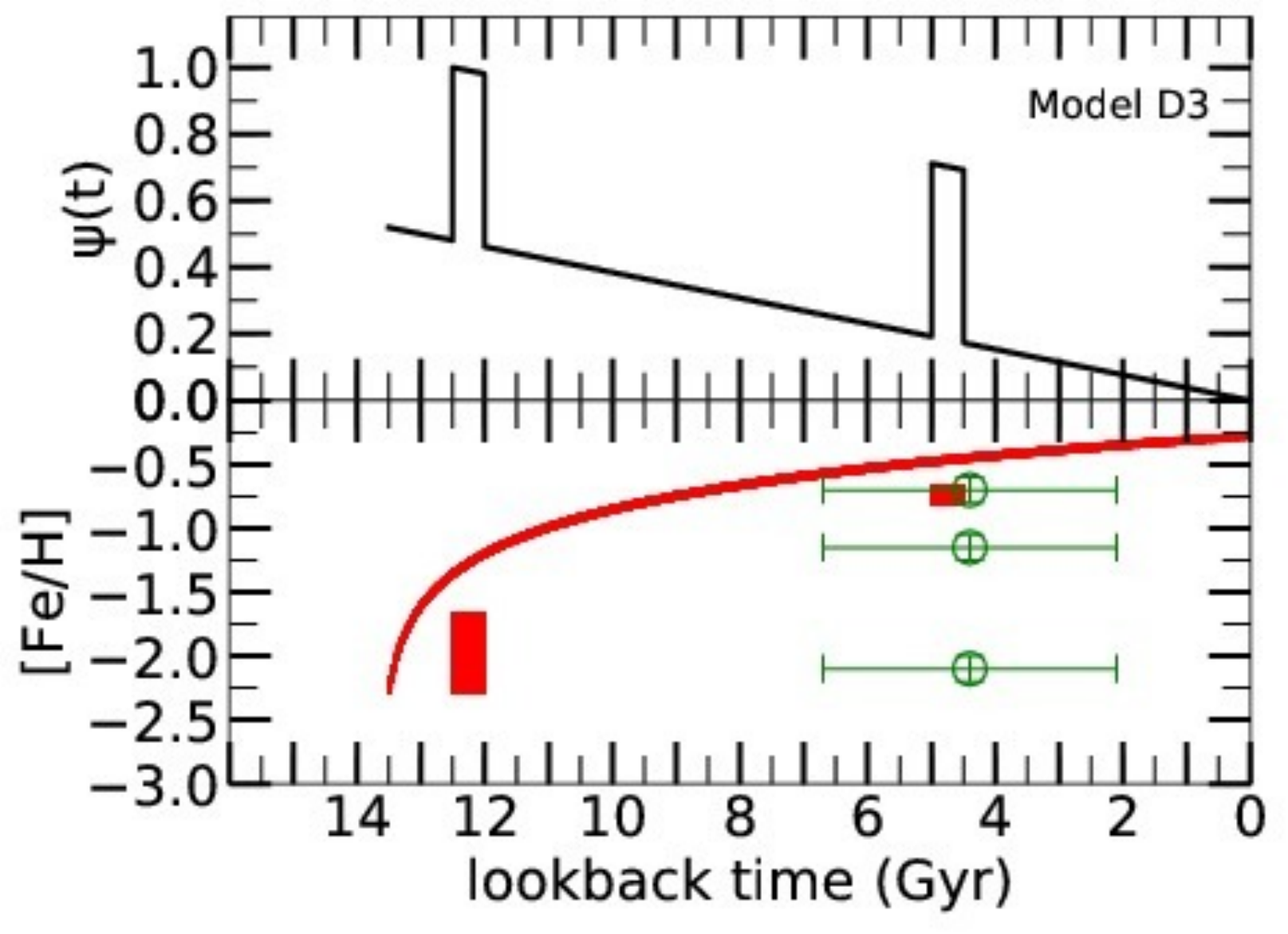}
\includegraphics[width=\columnwidth]{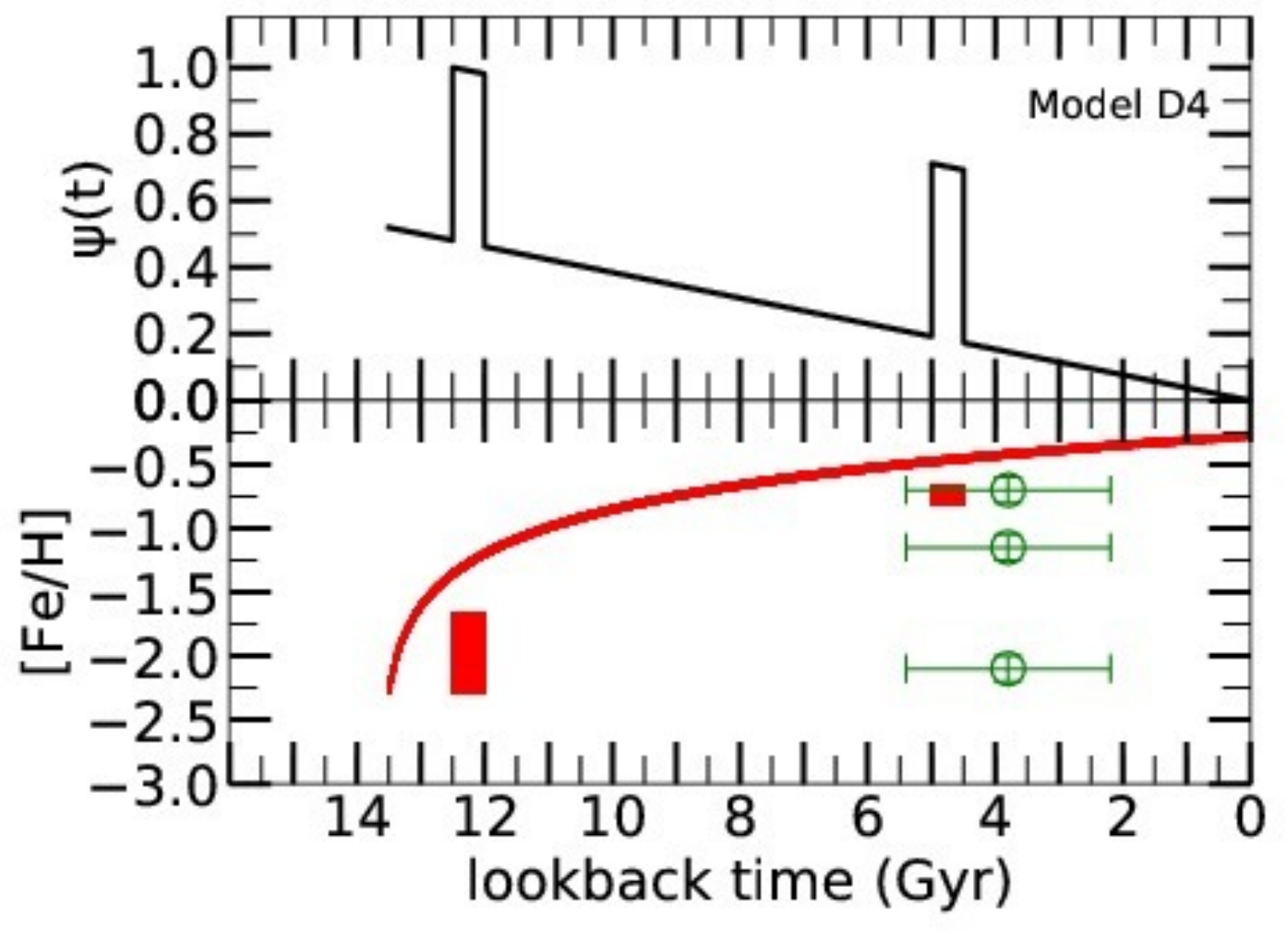}
    \caption{Same as Fig.~\ref{fig:fig2}, but for  stellar population model  D.}
   \label{fig:fig5}
\end{figure*}

\begin{figure*}
\includegraphics[width=\columnwidth]{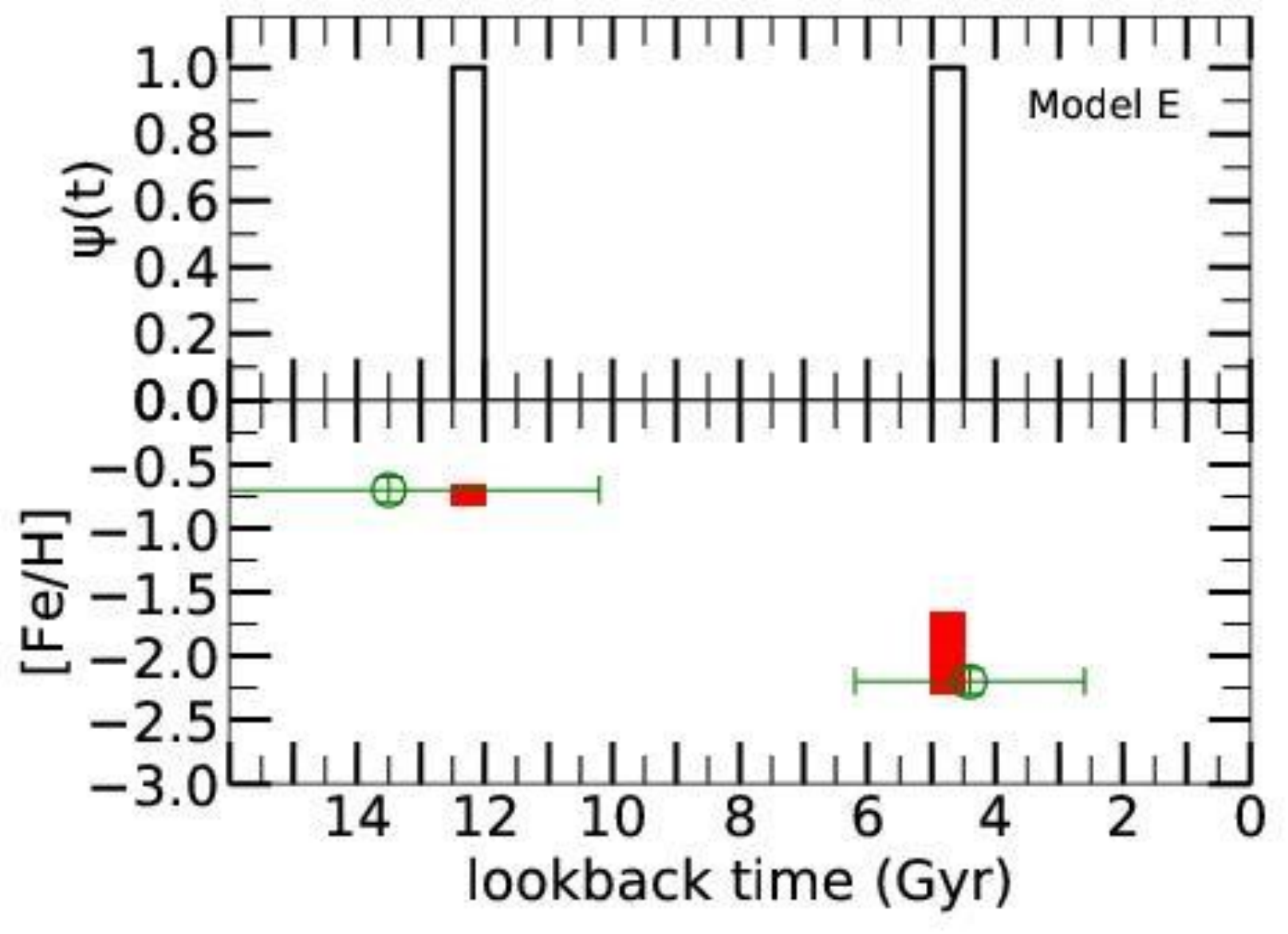}\\
\includegraphics[width=\columnwidth]{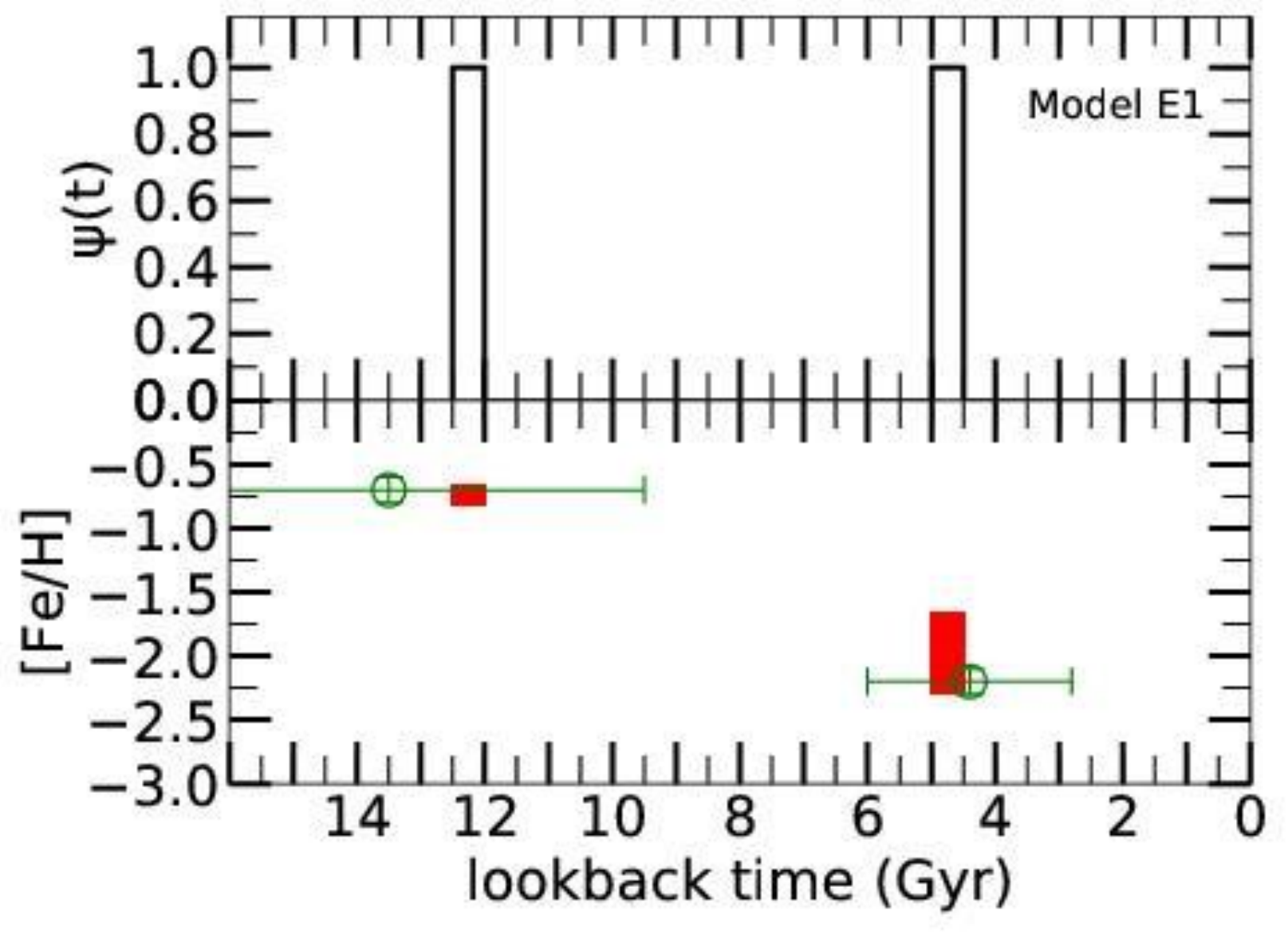}
\includegraphics[width=\columnwidth]{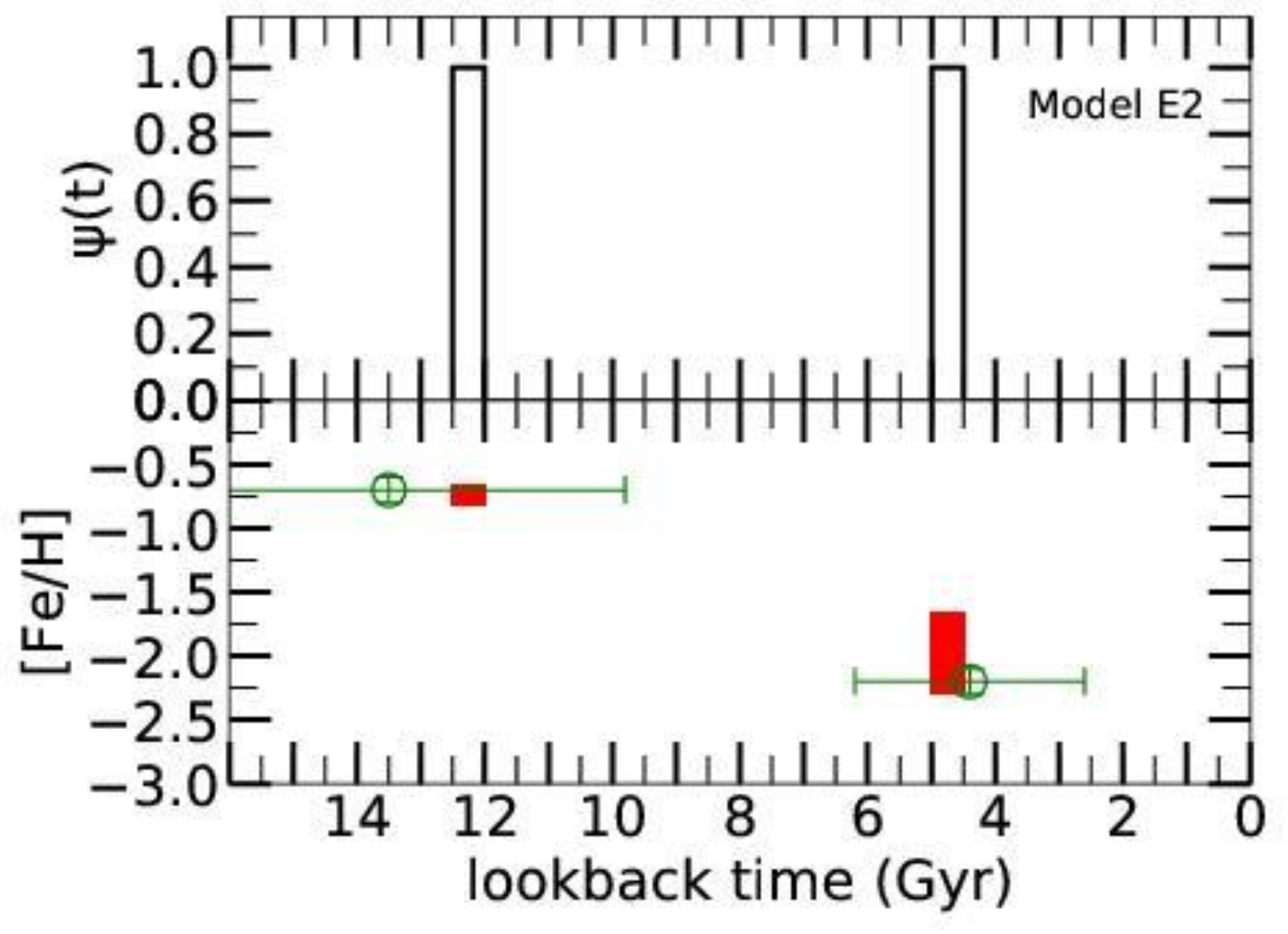}
\includegraphics[width=\columnwidth]{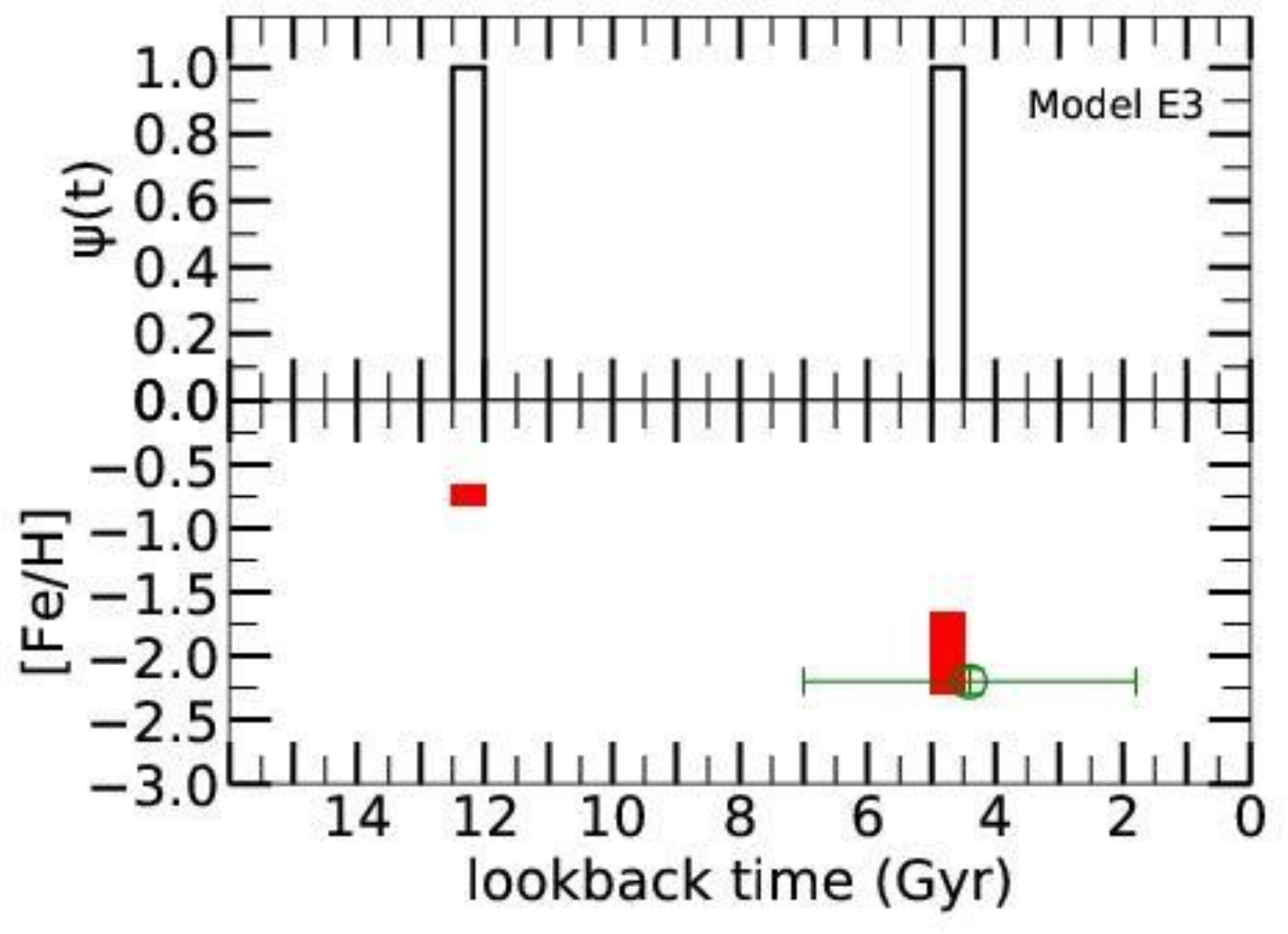}
\includegraphics[width=\columnwidth]{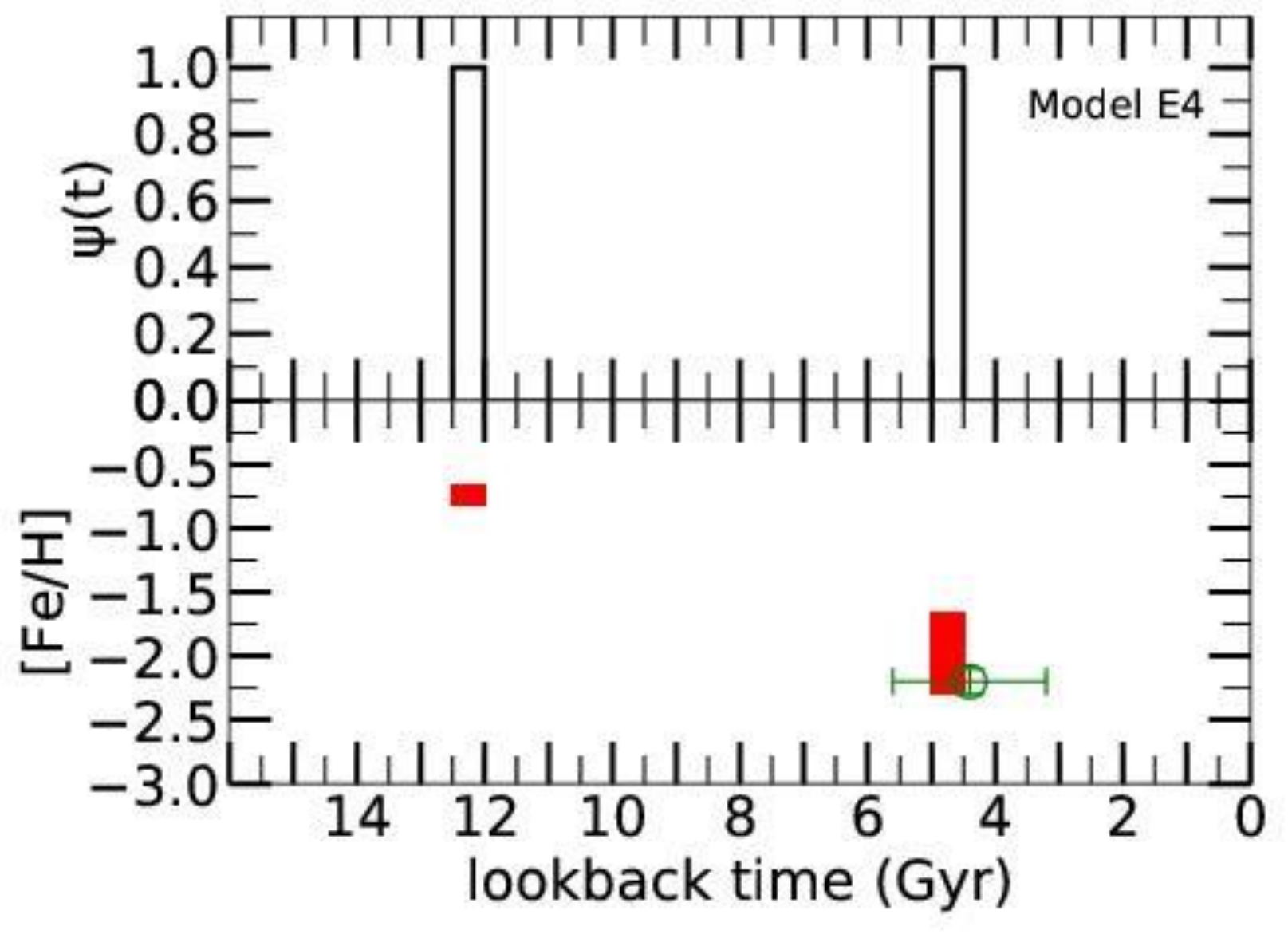}
    \caption{Same as Fig.~\ref{fig:fig2}, but for  stellar population model  E.}
   \label{fig:fig6}
\end{figure*}

We have used IAC-star \citep{ag04} to compute the four synthetic stellar populations. They have been labeled as A, B, C, D and E and their SFRs and AMRs are shown in  Figs.~\ref{fig:fig2} to ~\ref{fig:fig6}.
 The characteristics of the populations are as follows:
 population A consists in two bursts lasting 0.5 Gyr each and starting at ages 12.5 and 5 Gyr, respectively. A metallicity dispersion is used for each one: $0.0001\leq Z\leq 0.0004$ for the old one and $0.003\leq Z\leq 0.004$ for the young one. Population  B begins with a burst that starts at 12.5 Gyr in age, peaks at 12 Gyr and goes down to a low rate at 11 Gyr; after that, the star formation is low and goes on mildly decreasing to 0 at present. The metallicity follows a closed box model with initial value $Z_{i}=0.0004$, final value $Z_{f}=0.01$ and final gas mass fraction $\mu_{f}=0.1$; to this, a gaussian dispersion has been added such that $\sigma_Z/Z=0.25$. Population  C starts  at age 13.5 Gyr  with a high SFR that decreases  linearly to 0 at present. The metallicity increases linearly from an initial value of $Z_{i}=0.0001$ to a final value of $Z_{f}=0.01$. As in the former case, a gaussian metallicity dispersion has been added such that $\sigma_Z/Z=0.25$. Finally, population  D is simply the sum of populations  A and  C. In all the cases, a double power law has been assumed for the Initial Mass Function with exponent $x=-1.3$ for the stellar mass interval $0.1\leq m\leq 0.5$ M$_\odot$ and $x=-2.5$ for the interval $0.5\leq m\leq 120$ M$_\odot$. Also, a fraction of 30\% binary stars has been assumed, with a flat secondary to primary mass  ratio distribution with minimum value 0.5 and maximum, 1.  The number of stars in the $M_V$ vs. $(V-I)$ CMD, down to a magnitude limit of $M_I=5$, has been fixed in $10^5$ stars for models  A,  B and  C and $2\times 10^5$ stars for model  D. These figures translate in ever formed star masses of $\sim 1.3\times 10^7$ M$_\odot$ for models  A,  B and  C and $\sim 2.6\times 10^7$ M$_\odot$ for model  D.

\begin{figure}
	\includegraphics[width=\columnwidth]{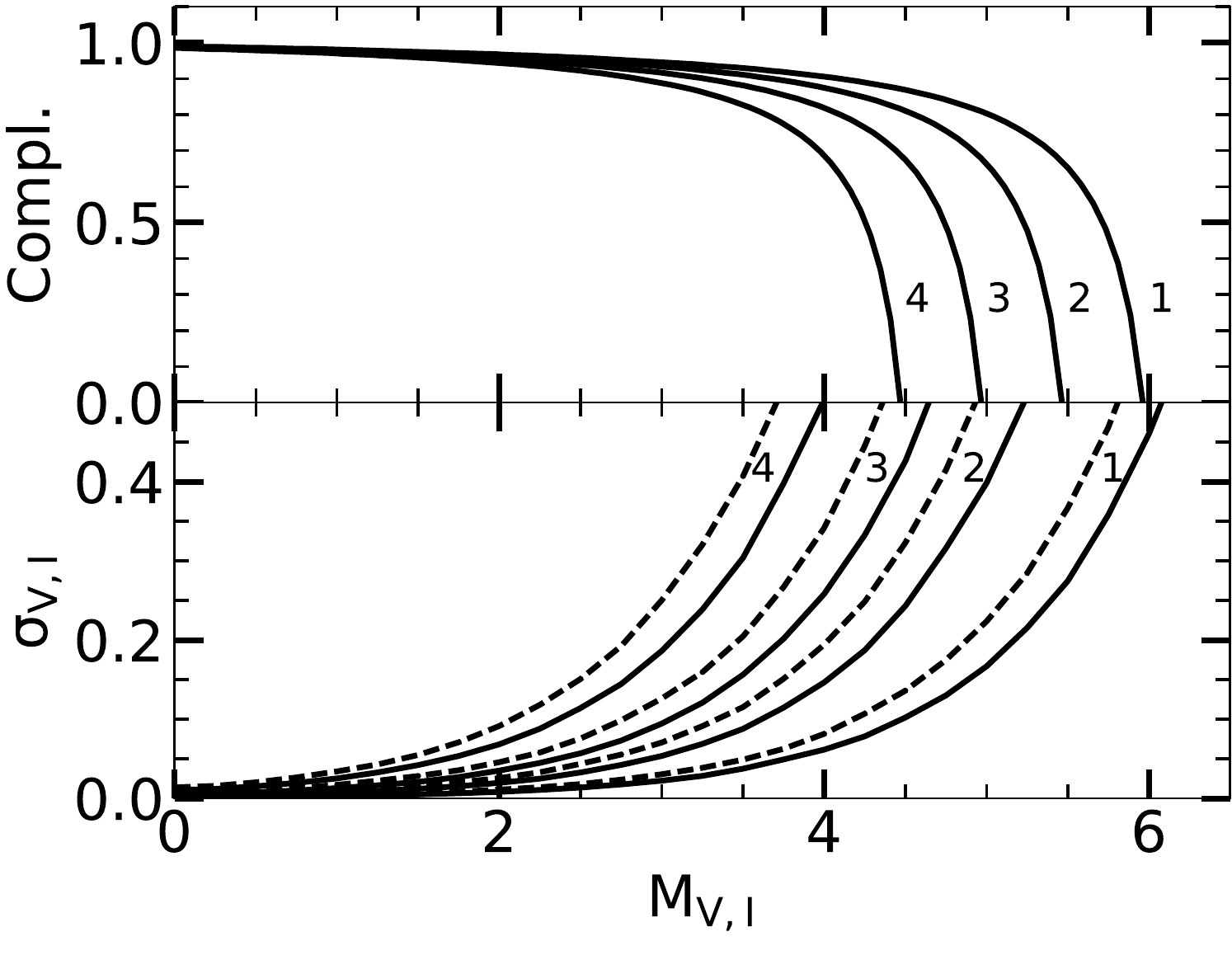}
    \caption{ Completeness fraction (upper panel) and errors (lower panel) for the four observational scenarios considered for each stellar population model. In the lower panel, full lines show $\sigma_V$ vs. $M_V$, while dashed lines refer to $\sigma_I$ vs. $M_I$.
    }
   \label{fig:fig7}
\end{figure}

\begin{figure*}
	\includegraphics[width=\columnwidth]{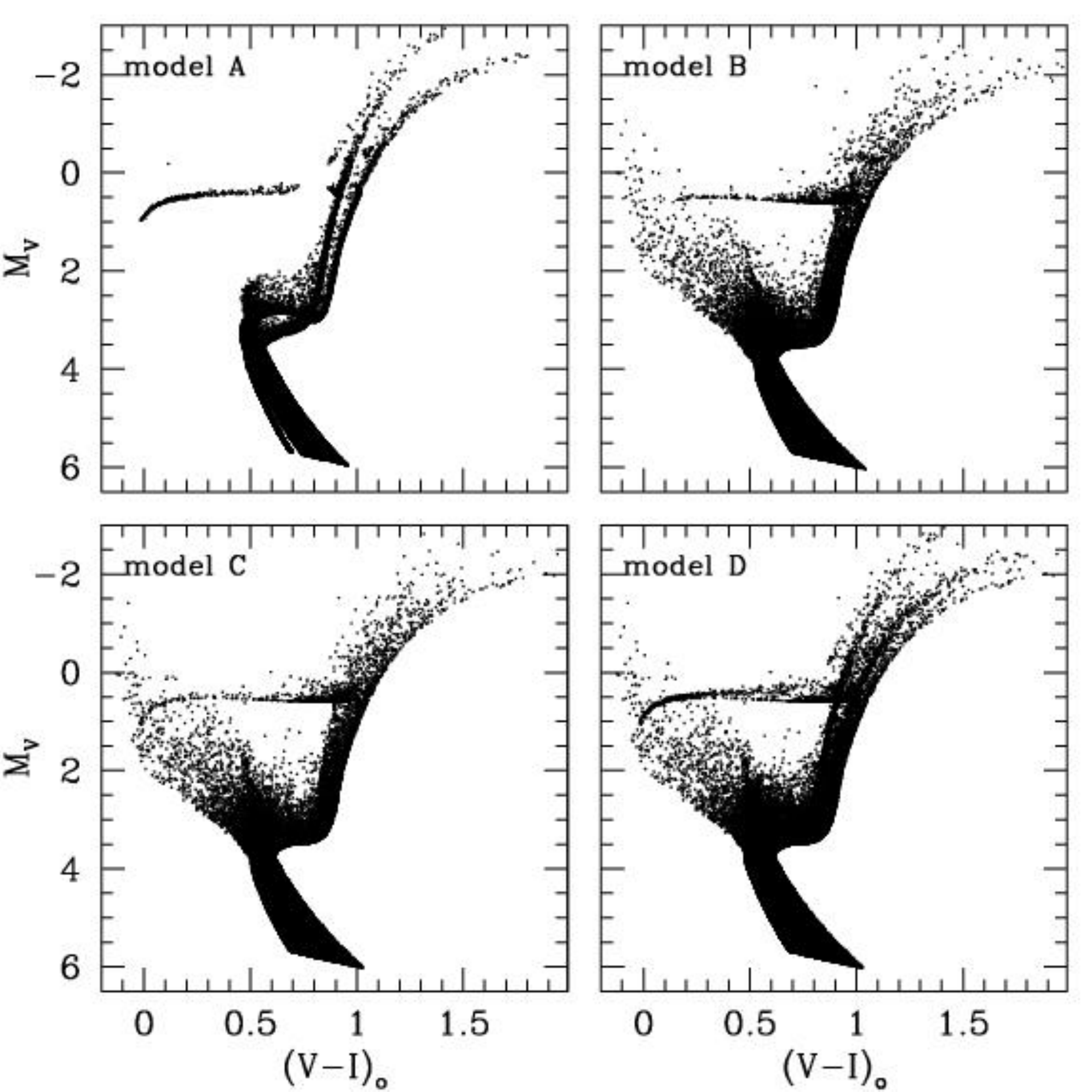}
	\includegraphics[width=\columnwidth]{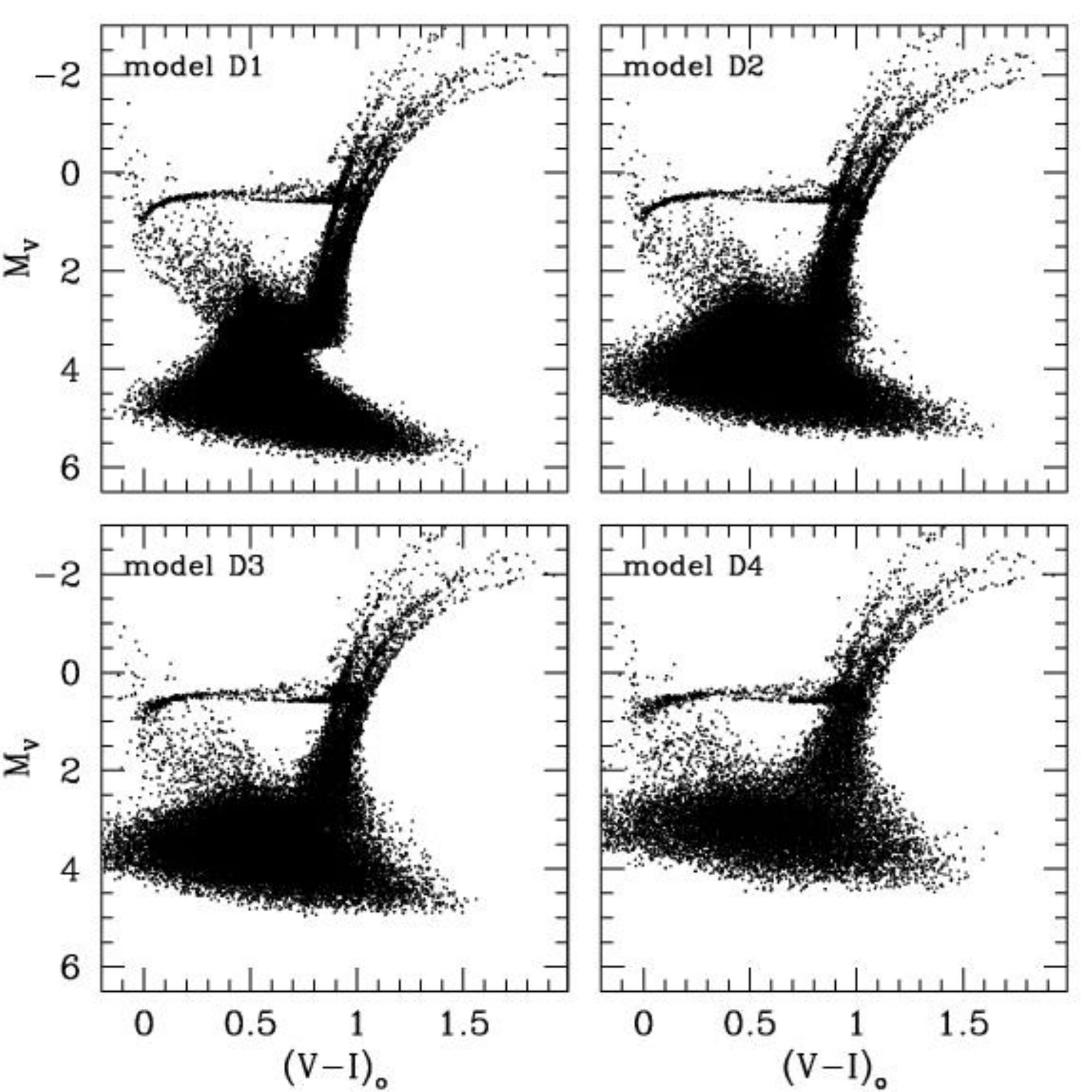}
    \caption{{\it Left:} Synthetic CMDs for models  A, B, C and D, without observational
effects. {\it Right:} CMDs for models  D for the four different scenarios with observation
effects taken into account (see text for details).}
   \label{fig:fig8}
\end{figure*}

Regarding observational effects,  four scenarios have been simulated in each population. The corresponding CMDs are labeled from 1 to 4 for each synthetic population; e.g.  A1, A2, A3,
and A4, for the case of population  A. Each scenario is represented by a completeness function and an error distribution. 
In Fig.~\ref{fig:fig7}, completeness as a function of $M_V$ (upper panel) and errors as a function of $M_V$ and $M_I$ (bottom panel) are plotted for the four scenarios. Completeness is simulated in the following way: for each star, the completeness fraction corresponding to its $M_V$ magnitude is obtained. Then a random number generator is used to maintain or eliminate the star from the photometry list; the star is conserved if the random number is smaller than the completeness fraction. Errors are simulated in the following way: for each conserved star, shifts in magnitudes in both filters are applied randomly selected from gaussian distributions which standard deviations are the values of $\sigma_V$ (plotted in Fig.~\ref{fig:fig7}) and $\sigma_I$, respectively. In order to illustrate the reader, Fig.~\ref{fig:fig8} depicts some of the generated synthetic CMDs.

\subsection{Representative AMRs}

\begin{figure}
	\includegraphics[width=\columnwidth]{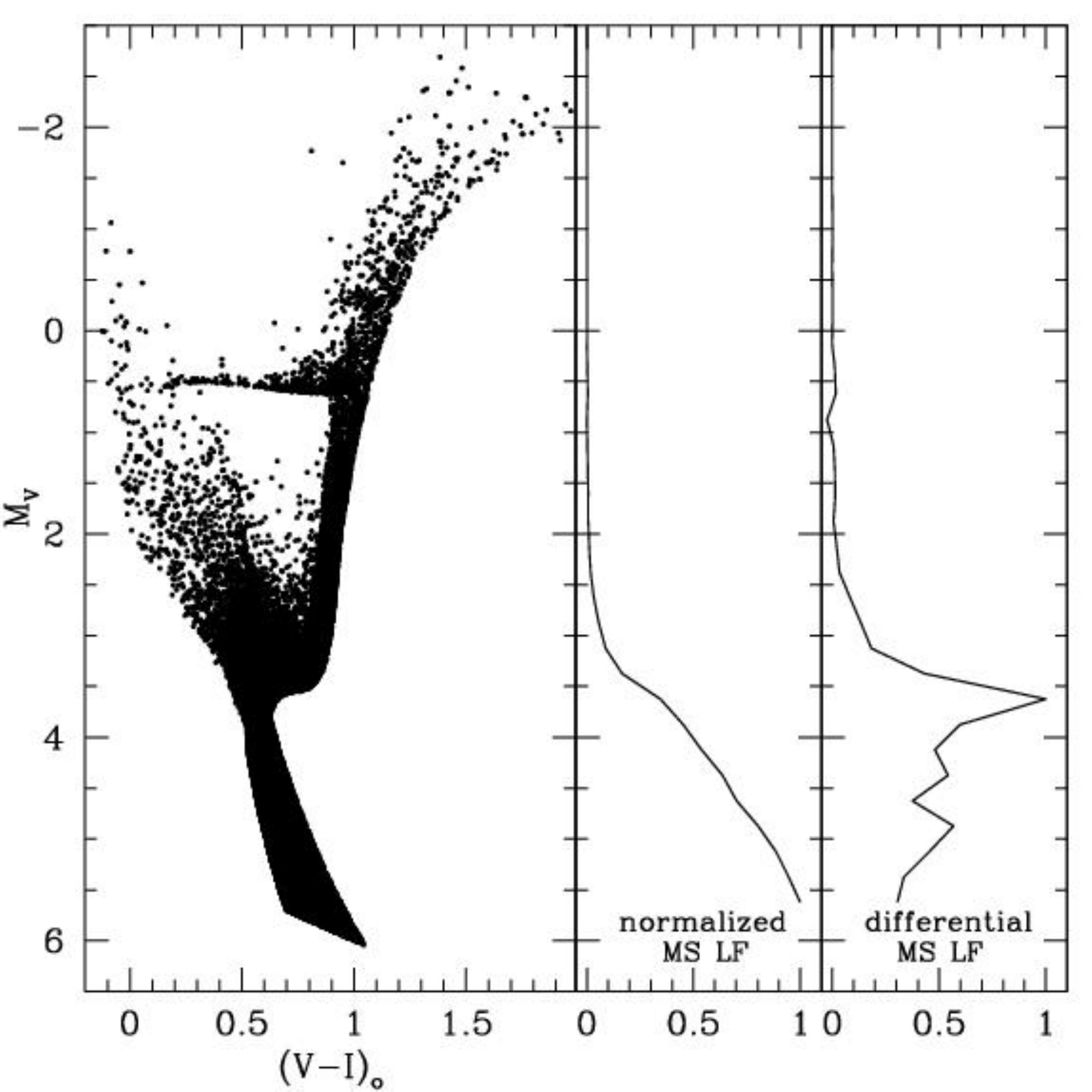}
    \caption{Synthetic $M_V,(V-I)_o$ diagram for model  B with the respective
normalized MS and differential MS LFs (see text for details).}
   \label{fig:fig9}
\end{figure}

Fig.~\ref{fig:fig9} illustrates, as an example, the generated $M_V$ vs $(V-I)_o$ CMD 
for model  B, which includes 100,000 stars.  A mixture of young through
old stellar populations clearly appears to be the main feature
of this CMD. Other obvious traits presented in the CMD are the populous and broad
subgiant branches  (an indicator of the evolution of stars with
ages (masses) within a non-negligible range), the RCs and the
RGBs. In the middle panel of  Fig.~\ref{fig:fig9} we drew the respective MS LF. 
The whole set of MS LFs (one for each model) 
were obtained by counting the number of stars in $M_V$ bins of 0.25 mag along the MSs, 
and then they were normalized. The chosen bin size encompasses typical magnitude errors 
of the stars in each bin, thus producing an appropriate sample of the stars. Note that
accurate photometries have magnitude errors for most of the measured stars usually
smaller than 0.20 mag. As is well-known, the bin size should be of
the order of the uncertainties of the quantity involved to best
represent an intrinsic distribution of such a quantity  \citep{p10,p11a}.
This means $-$ statistically speaking $-$ that the shape of these LFs
are not driven by the chosen bin size.

We then computed for each synthetic CMD the difference between the number of stars
of two adjacent magnitude intervals to produce the  differential MS LF,
as illustrated in the right panel of   Fig.~\ref{fig:fig9}.  As described in
Section 2, the peak and the broadness of the differential MS LF provide
information about the representative TO magnitude and its  intrinsic
width of the considered stellar populations.
The prevailing  TO (differential MS LF peak) of each synthetic CMD resulted typically 
$\sim 25\%-50\%$ more populous than the next most dominant population,
represented by a secondary peak $-$ sometimes there even could exist a third peak $-$ in
the differential LFs. We adopted here  the two main peaks we distinguished in the 
differential MS LFs, as well as their observed FWHMs, as a measure of their
intrinsic broadnesses.  For the synthetic CMDs of models A, D, D1, D2, D3 and D4 we 
distinguished three main peaks, whereas for those of models A3 and A4 only one peak. 
Table~\ref{tab:table2} lists the
representative $M_V$(TO) with their associated widths for each modelled SFH.

As for the representative RCs, we  built $M_V$ distributions for the RC stars 
and performed gaussian fits to derive the mean values and the FWHMs.
We performed gaussian fits using the {\sc ngaussfit} routine of the
IRAF\footnote{IRAF is distributed by the National 
Optical Astronomy Observatories, which is operated by the Association of 
Universities for Research in Astronomy, Inc., under contract with the National 
Science Foundation.} {\sc stsdas} package. We adopted a single  gaussian, and fixed
the constant and linear terms to the corresponding background
levels and to zero, respectively. The centre of the gaussian, its
amplitude, and its FWHM acted as variables. Table~\ref{tab:table2} lists the
resulting representative $M_V$(RC).

Since we are primarily interested in determining the age and
metallicity of the representative star population in each model,
we derived $\delta V$ indices by calculating the
difference in the $M_V$ magnitude between the RC and the MSTO. 
We then derived ages from the $\delta V$ values by using the
equation \citep{petal14c}:

\begin{equation}
age(Gyr) = 0.538 + 1.795\delta V - 1.480(\delta V)^2 + 0.626(\delta V)^3
\end{equation}

This equation is only calibrated for ages
larger than 1 Gyr, so that we are not able to produce ages
for younger representative populations. Table~\ref{tab:table2} presents the
resultant ages and their dispersions. The dispersions have
been calculated bearing in mind the broadness of the $M_V$ mag
distributions of the representative MSTOs and RCs, and represent in general 
a satisfactory estimate of the age spread around the prevailing population
age.

In addition, we also estimated representative metallicities using the equation:

\begin{equation}
{\rm [Fe/H]} = -15.16 + 17.0(V-I)_{o,-3} - 4.9(V-I)^2_{o,-3}
\end{equation} 

\noindent of \citet{da90}, once the $(V-I)_o$ colours of the
RGB at $M_I$ =  $-$3.0 mag and their dispersions were obtained. The $(V-I)_o$ colours 
were derived from the intersection of the RGBs traced for each model and the 
horizontal line at $M_I$ =  $-$3.0 mag. Table~\ref{tab:table2}  provides with the
resulting values and the derived representative metallicities.

\begin{table*}
\caption{Representative parameter values for the generated models.}
\label{tab:table2}
\begin{tabular}{@{}lccccc}\hline
Model & $M_V$(TO) & $M_V$(RC) & age & $(V-I)_{o,-3}$ & [Fe/H] \\
      &  (mag)    &  (mag)    & (Gyr) &  (mag) & (dex) \\\hline
 A     & 2.875$\pm$0.200 & 0.450$\pm$0.020 &  5.1$\pm$1.2 & 1.35$\pm$0.02 &  $-$1.15$\pm$0.1\\
      & 3.500$\pm$0.500 & 0.450$\pm$0.020 & 10.0$\pm$5.4 & 1.15$\pm$0.02 &  $-$2.1$\pm$0.1\\
      & 4.875$\pm$0.400 & 0.450$\pm$0.020 & 13.5$^a$     & 1.15$\pm$0.02 &  $-$2.1$\pm$0.1\\
 A1    & 2.875$\pm$0.200 & 0.450$\pm$0.020 &  5.1$\pm$1.2 & 1.35$\pm$0.02 &  $-$1.15$\pm$0.1\\
      & 3.375$\pm$0.400 & 0.450$\pm$0.020 &  8.8$\pm$3.9 & 1.15$\pm$0.02 &  $-$2.1$\pm$0.1\\
 A2    & 2.875$\pm$0.250 & 0.450$\pm$0.020 &  5.1$\pm$1.4 & 1.35$\pm$0.02 &  $-$1.15$\pm$0.1\\
      & 3.375$\pm$0.300 & 0.450$\pm$0.020 &  8.8$\pm$3.9 & 1.15$\pm$0.02 &  $-$2.1$\pm$0.1\\
 A3    & 2.875$\pm$0.250 & 0.450$\pm$0.020 &  5.1$\pm$1.4 & 1.35$\pm$0.02 &  $-$1.15$\pm$0.1\\
 A4    & 2.750$\pm$0.300 & 0.450$\pm$0.020 &  4.5$\pm$1.5 & 1.35$\pm$0.02 &  $-$1.15$\pm$0.1\\
 B     & 3.625$\pm$0.350 & 0.600$\pm$0.020 &  9.7$\pm$3.7 & 1.55$\pm$0.02 &  $-$0.6$\pm$0.1\\
      & 4.875$\pm$0.300 & 0.600$\pm$0.020 & 13.5$^a$     & 1.25$\pm$0.02 &  $-$1.6$\pm$0.1\\
B1    & 3.750$\pm$0.600 & 0.600$\pm$0.020 & 11.1$\pm$6.0 & 1.55$\pm$0.02 &  $-$0.6$\pm$0.1\\
      & 3.750$\pm$0.600 & 0.600$\pm$0.020 & 11.1$\pm$6.0 & 1.25$\pm$0.02 &  $-$1.6$\pm$0.1\\
 B2   & 3.750$\pm$0.300 & 0.600$\pm$0.020 & 11.1$\pm$3.0 & 1.55$\pm$0.02 &  $-$0.6$\pm$0.1\\
      & 3.750$\pm$0.300 & 0.600$\pm$0.020 & 11.1$\pm$3.0 & 1.25$\pm$0.02 &  $-$1.6$\pm$0.1\\
 B3    & 3.250$\pm$0.450 & 0.600$\pm$0.020 &  6.6$\pm$4.0 & 1.55$\pm$0.02 &  $-$0.6$\pm$0.1\\
      & 3.250$\pm$0.450 & 0.600$\pm$0.020 &  6.6$\pm$4.0 & 1.25$\pm$0.02 &  $-$1.6$\pm$0.1\\
 B4    & 2.875$\pm$0.300 & 0.600$\pm$0.020 &  4.5$\pm$1.5 & 1.55$\pm$0.02 &  $-$0.6$\pm$0.1\\
      & 2.875$\pm$0.300 & 0.600$\pm$0.020 &  4.5$\pm$1.5 & 1.25$\pm$0.02 &  $-$1.6$\pm$0.1\\
 C     & 3.675$\pm$0.300 & 0.590$\pm$0.020 & 10.4$\pm$3.4 & 1.50$\pm$0.02 &  $-$0.7$\pm$0.1\\
       & 3.675$\pm$0.300 & 0.590$\pm$0.020 & 10.4$\pm$3.4 & 1.35$\pm$0.02 &  $-$1.15$\pm$0.1\\
 C1    & 3.675$\pm$0.450 & 0.590$\pm$0.020 & 10.4$\pm$5.0 & 1.50$\pm$0.02 &  $-$0.7$\pm$0.1\\
      & 3.675$\pm$0.450 & 0.590$\pm$0.020 & 10.4$\pm$5.0 & 1.35$\pm$0.02 &  $-$1.15$\pm$0.1\\
 C2    & 3.375$\pm$0.300 & 0.590$\pm$0.020 &  7.6$\pm$3.0 & 1.50$\pm$0.02 &  $-$0.7$\pm$0.1\\
      & 3.375$\pm$0.300 & 0.590$\pm$0.020 &  7.6$\pm$3.0 & 1.35$\pm$0.02 &  $-$1.15$\pm$0.1\\
 C3    & 3.375$\pm$0.400 & 0.590$\pm$0.020 &  7.6$\pm$3.5 & 1.50$\pm$0.02 &  $-$0.7$\pm$0.1\\
      & 3.375$\pm$0.400 & 0.590$\pm$0.020 &  7.6$\pm$3.5 & 1.35$\pm$0.02 &  $-$1.15$\pm$0.1\\
 C4    & 2.675$\pm$0.350 & 0.590$\pm$0.020 &  3.5$\pm$1.4 & 1.50$\pm$0.02 &  $-$0.7$\pm$0.1\\
      & 2.675$\pm$0.350 & 0.590$\pm$0.020 &  3.5$\pm$1.4 & 1.35$\pm$0.02 &  $-$1.15$\pm$0.1\\
 D    & 2.875$\pm$0.200 & 0.590$\pm$0.020 &  4.4$\pm$1.1 & 1.50$\pm$0.02 &  $-$0.7$\pm$0.1\\
      & 3.625$\pm$0.500 & 0.590$\pm$0.020 &  9.85$\pm$5.3& 1.35$\pm$0.02 &  $-$1.15$\pm$0.1\\
      & 4.875$\pm$0.400 & 0.590$\pm$0.020 & 13.5$^a$     & 1.15$\pm$0.02 &  $-$2.1$\pm$0.1\\
 D1    & 2.875$\pm$0.200 & 0.590$\pm$0.020 &  4.4$\pm$1.1 & 1.50$\pm$0.02 &  $-$0.7$\pm$0.1\\
      & 3.625$\pm$0.700 & 0.590$\pm$0.020 &  9.85$\pm$7.5& 1.35$\pm$0.02 &  $-$1.15$\pm$0.1\\
      & 3.625$\pm$0.700 & 0.590$\pm$0.020 &  9.85$\pm$7.5& 1.15$\pm$0.02 &  $-$2.1$\pm$0.1\\
 D2    & 2.875$\pm$0.200 & 0.590$\pm$0.020 &  4.4$\pm$1.1 & 1.50$\pm$0.02 &  $-$0.7$\pm$0.1\\
      & 3.375$\pm$0.350 & 0.590$\pm$0.020 &  7.6$\pm$3.0 & 1.35$\pm$0.02 &  $-$1.15$\pm$0.1\\
      & 3.375$\pm$0.350 & 0.590$\pm$0.020 &  7.6$\pm$3.0 & 1.15$\pm$0.02 &  $-$2.1$\pm$0.1\\
 D3    & 2.875$\pm$0.450 & 0.590$\pm$0.020 &  4.4$\pm$2.3 & 1.50$\pm$0.02 &  $-$0.7$\pm$0.1\\
      & 2.875$\pm$0.450 & 0.590$\pm$0.020 &  4.4$\pm$2.3 & 1.35$\pm$0.02 &  $-$1.15$\pm$0.1\\
      & 2.875$\pm$0.450 & 0.590$\pm$0.020 &  4.4$\pm$2.3 & 1.15$\pm$0.02 &  $-$2.1$\pm$0.1\\
 D4    & 2.750$\pm$0.350 & 0.590$\pm$0.020 &  3.8$\pm$1.6 & 1.50$\pm$0.02 &  $-$0.7$\pm$0.1\\
      & 2.750$\pm$0.350 & 0.590$\pm$0.020 &  3.8$\pm$1.6 & 1.35$\pm$0.02 &  $-$1.15$\pm$0.1\\
      & 2.750$\pm$0.350 & 0.590$\pm$0.020 &  3.8$\pm$1.6 & 1.15$\pm$0.02 &  $-$2.1$\pm$0.1\\
 E & 2.500$\pm$0.250 & 0.200$\pm$0.020 &  4.4$\pm$1.8 & 1.50$\pm$0.02 &  $-$0.7$\pm$0.1\\
       & 3.750$\pm$0.200 & 0.200$\pm$0.020 & 13.5$\pm$3.3 & 1.10$\pm$0.02 &  $-$2.2$\pm$0.1\\
    E1 & 2.500$\pm$0.300 & 0.200$\pm$0.020 &  4.4$\pm$1.6 & 1.50$\pm$0.02 &  $-$0.7$\pm$0.1\\
       & 3.750$\pm$0.250 & 0.200$\pm$0.020 & 13.5$\pm$4.0 & 1.10$\pm$0.02 &  $-$2.2$\pm$0.1\\
  E2 & 2.500$\pm$0.350 & 0.200$\pm$0.020 &  4.4$\pm$1.8 & 1.50$\pm$0.02 &  $-$0.7$\pm$0.1\\
       & 3.625$\pm$0.250 & 0.200$\pm$0.020 & 13.5$\pm$3.7 & 1.10$\pm$0.02 &  $-$2.2$\pm$0.1\\
E3 & 2.500$\pm$0.500 & 0.200$\pm$0.020 &  4.4$\pm$2.6 &1.50$\pm$0.02 &  $-$0.7$\pm$0.1\\
E4 & 2.500$\pm$0.250 & 0.200$\pm$0.020 &  4.4$\pm$1.2 & 1.50$\pm$0.02 &  $-$0.7$\pm$0.1\\
\hline
\end{tabular}

\noindent $^a$ adopted value due to $\delta V$ falls outside the range of eq. (1).
\end{table*}

\section{Discussion and Conclusions}

In the bottom panels of  Figures~\ref{fig:fig2} to  \ref{fig:fig6} we superimposed the representative ages and metallicities derived in Sect. 3.2  (green open boxes)
to the modelled AMRs (red lines). The errorbars represent the intrinsic FWHMs of such 
representative values.  In this Section we compare the trends suggested by the representative age/metallicity values with those coming directly from the modelled SFHs. In other words,
we assess how well main changes in the modelled AMRs are detected by  the representative
AMRs. Note that the representative method is not meant to reproduce any particular
modelled stellar population, but to map the main trends in the present-day AMR from the so-called
representative populations.

At first glance,  there are clear differences in the success with which the 
representative AMRs represent the actual AMR. For instance,  completeness effects 
 constrain significantly the performance of representative AMRs, a result which is expected since 
representative AMRs come from the employment of observed CMDs with different completeness 
effects.  From the inspection of Figures~\ref{fig:fig2} to \ref{fig:fig6} we found that the oldest representative age is highly dependent  on this completeness
factor. As can be seen, the lower the photometry completeness the younger the oldest representative age derived. They are in agreement (considering their errorbars) with
the oldest modelled ages only for models without completeness effect (models A, B, C,  D and E) and 
models with completeness effect \# 1 and 2, and in some few cases \# 3. Thus, by entering into
 Fig.~\ref{fig:fig7}
with the absolute magnitude $M_V$ associated to the MSTO of  these oldest representative ages
(see Table~\ref{tab:table2}) we found that  the associated completeness factor is higher 
than $\sim$ 85 per cent. This means that reliable oldest representative ages can be obtained
for stellar populations with a photometry completeness higher than $\sim$ 85 per cent. Dominant
stellar populations observed with less complete photometry do not come up as representative ones.
This is somehow an expected  result, since it would be hardly possible to say anything about
a prevailing stellar population if it were not observed mostly complete.
Indeed, the representative AMRs built by \citet{p12a,pg13} and
\citet{petal14c} for the SMC, LMC and the Fornax dwarf spheroidal, respectively,
were obtained from photometry with completeness factors higher than  90 per cent, and 
hence the good agreement seen with other independent AMRs.

As for the metallicities, we found that the representative AMRs were able to 
 reasonably account for the modelled  metallicity 
enrichment ranges when they  arose as a consequence of a bursting formation event 
(models A, B, D and E). Particularly, the metal-poor ends were mostly well reproduced, while the
metal-rich one resulted $\sim$ 0.3--0.4 dex more metal-poor. The latter could be due to
a relatively low metallicity  sensitivity of the employed photometric system toward the
metal-rich end. This does not happen, for instance, for the Washington photometric system
which is more sensitive to metallicity (see. Fig.~\ref{fig:fig1}). Likewise, the modelled age
ranges were more  satisfactorily reproduced by the resulting representative age ranges whenever
models contained bursting formation events.

During quiescent periods of star formation or periods with a decreasing SFR,
as is the case of model C, the
representative AMRs provide only with mean values of age and metallicity for all
the stars formed  at $\Psi$(t) higher than $\sim$ 0.6. This can be easily checked
by entering into Fig.~\ref{fig:fig4} with the representative ages (mean values and errors)
and interpolate the $\Psi$(t) one for model C (without completeness effect).
This means that  detected representative populations consist of
any stellar population whose stars  are within at least $\sim$ 40 per cent of the
most massive one in the galaxy. 
As we mention in Sect. 2, the definition of a representative TO 
-- and hence of a representative stellar population -- could not converge
to any dominant TO (age) value if the stars in a given field came
from a constant star formation rate (SFR) integrated over all time. Likewise, minority 
stellar populations not following these main chemical galactic processes are discarded. 
Both nearly constant SFR and minority population effects can be seen in Fig.~\ref{fig:fig3} 
 to  \ref{fig:fig6} (models B, C,  D and E), where only main bursts were detected. 

On the other hand, we confirm that the representative method is suitable to detect mayor
star formation events in the galaxy lifetime, such as bursts of stellar populations. For instance, bursts in models  A (at $\sim$ 5 and 12 Gyr), B (at $\sim$ 12 Gyr),
 D (at $\sim$ 
5 and 12 Gyr)  and   E (at $\sim$ 
5 and 12 Gyr) are well detected, provided a good photometric completeness. This is because 
a mayor bursting event can produce a significant amount of stars (not necessarily followed
by a chemical enrichment), and therefore representative populations during the galaxy 
lifetime can emerge.  Moreover, even though the B, C and D modelled AMRs are similar  
-- they consist of an important increase in the metallicity at the very begining of the galaxy formation and nearly flat curves with a small slope around [Fe/H] $\sim$ $-$0.5 dex over 
most of the galaxy lifetime  -- the representative AMRs resulted different because of the 
differences in the SFHs. Those representative AMRs obtained from models with bursting 
stellar formation events reproduced better the age/metallicty modelled ranges.
Successive bursts of star formation could be
recognized if the conditions mentioned above
about the completeness factor and the $\Psi$(t) value are  fulfilled.

\begin{figure}
	\includegraphics[width=\columnwidth]{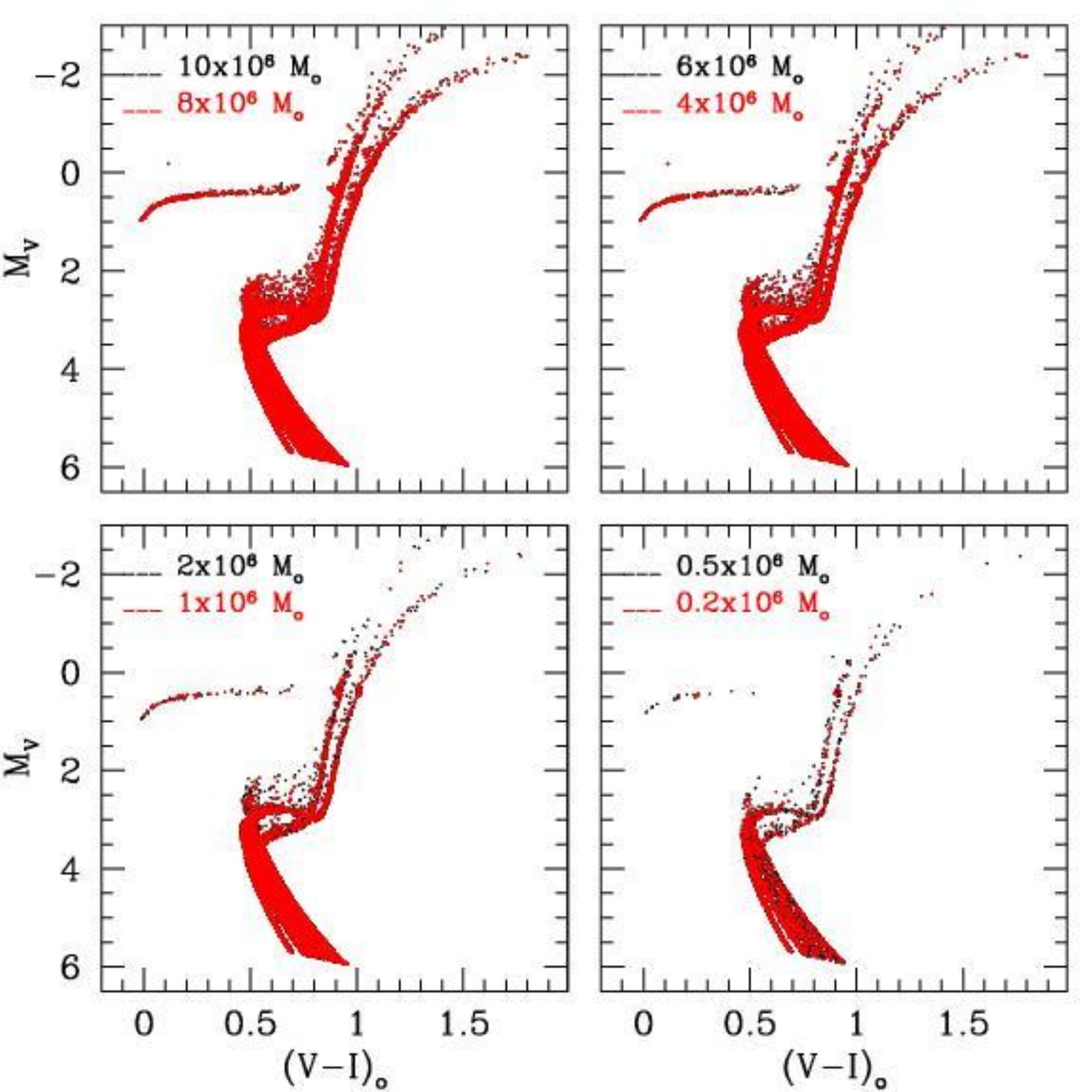}
    \caption{Synthetic $M_V,(V-I)_o$ diagrams for model  A for different
total galaxy mass, as indicated in each panel.}
   \label{fig:fig10}
\end{figure}

\begin{figure}
	\includegraphics[width=\columnwidth]{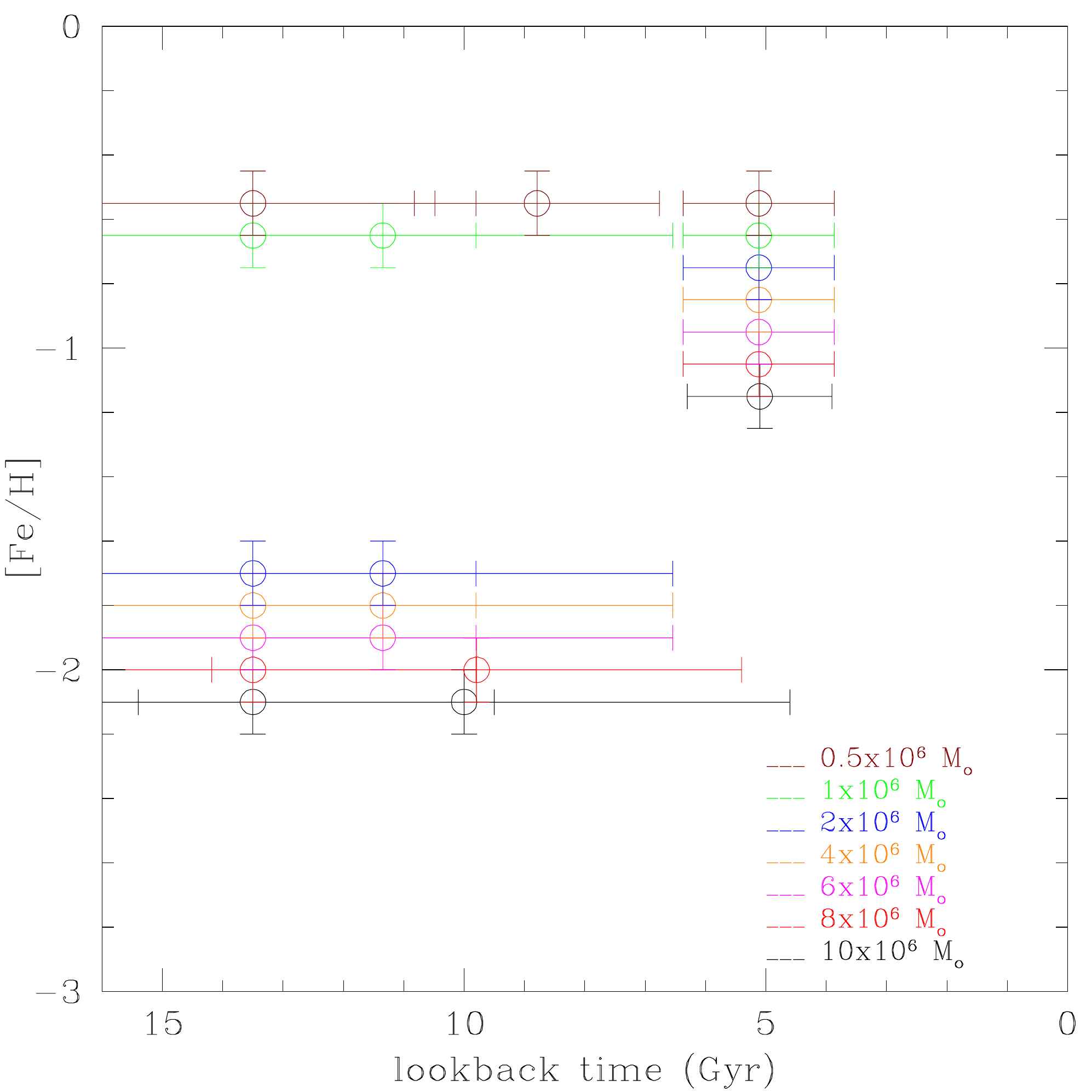}
    \caption{Representative AMRs for model  A for different
total stellar galaxy mass. We have shifted the points in [Fe/H] to avoid superposition. }
   \label{fig:fig11}
\end{figure}

Recovering the dominant stellar population(s), and hence the representative AMRs, 
depends on the total mass of the galaxy. This is because a representative AMR relies on 
the composite CMD of the galaxy stellar populations. If  real representative MSTOs,
RCs or RGBs are not visible in the CMD, the method
cannot be employed. To illustrate this point we generated
synthetic stellar populations (synthetic CMDs) with different input
total masses for our model A. The resulting synthetic CMDs are shown in Fig.~\ref{fig:fig10}.
As can be seen, below $\sim$ 1-2$\times$10$^6$ M$_\odot$, the redder RGB and the brighter MSTO are
difficult to recognize.  We then derived representative ages and metallicities for each
model and drew Fig.~\ref{fig:fig11} which compares the different outcomes. Once again, the originally
modelled AMR (model A) is not recovered for galaxy masses smaller than 1-2$\times$10$^6$ M$_\odot$.
For galaxy masses smaller than 0.5$\times$10$^6$ M$_\odot$, the RGB was not detected, so that
we could not estimate representative metallicities.

The ongoing surveys of galaxies, e.g., \citet[][CSI]{kelsonetal2014}, \citet[][LEGUS]{calzettietal2015} and those
for the next generation of telescopes will demand plenty of CPU time to recover a detailed 
galaxy SFH. This challenge points to the need of expeditive methods for obtaining
quick-look AMRs before  high-CPU consuming machine codes can be fully executed. 
The representative method presented here could be of a great help  in this respect, providing
AMRs  in advance (nearly in real time respect to the availability of observational data)
which statistically reflect the most important trends in the galaxy formation and
chemical evolution. The representative AMRs turn out to be reliable down to a magnitude limit with
a photometric completeness factor higher than $\sim$ 85 per cent, and trace the
chemical evolution history for any stellar population (represented by a mean age
and an intrinsic age spread) with a total mass  within  $\sim$ 40 per cent of
 the most massive stellar population in the galaxy.

\section*{Acknowledgements}
 We thank the referee, Jacco van Loon, for his thorough reading of the manuscript and
timely suggestions to improve it.




\bibliographystyle{mnras}

\begin{thebibliography}{}
\makeatletter
\relax
\def\mn@urlcharsother{\let\do\@makeother \do\$\do\&\do\#\do\^\do\_\do\%\do\~}
\def\mn@doi{\begingroup\mn@urlcharsother \@ifnextchar [ {\mn@doi@}
  {\mn@doi@[]}}
\def\mn@doi@[#1]#2{\def\@tempa{#1}\ifx\@tempa\@empty \href
  {http://dx.doi.org/#2} {doi:#2}\else \href {http://dx.doi.org/#2} {#1}\fi
  \endgroup}
\def\mn@eprint#1#2{\mn@eprint@#1:#2::\@nil}
\def\mn@eprint@arXiv#1{\href {http://arxiv.org/abs/#1} {{\tt arXiv:#1}}}
\def\mn@eprint@dblp#1{\href {http://dblp.uni-trier.de/rec/bibtex/#1.xml}
  {dblp:#1}}
\def\mn@eprint@#1:#2:#3:#4\@nil{\def\@tempa {#1}\def\@tempb {#2}\def\@tempc
  {#3}\ifx \@tempc \@empty \let \@tempc \@tempb \let \@tempb \@tempa \fi \ifx
  \@tempb \@empty \def\@tempb {arXiv}\fi \@ifundefined
  {mn@eprint@\@tempb}{\@tempb:\@tempc}{\expandafter \expandafter \csname
  mn@eprint@\@tempb\endcsname \expandafter{\@tempc}}}

\bibitem[\protect\citeauthoryear{{Aparicio} \& {Gallart}}{{Aparicio} \&
  {Gallart}}{2004}]{ag04}
{Aparicio} A.,  {Gallart} C.,  2004, \mn@doi [\aj] {10.1086/382836}, \href
  {http://adsabs.harvard.edu/abs/2004AJ....128.1465A} {128, 1465}

\bibitem[\protect\citeauthoryear{{Aparicio} \& {Hidalgo}}{{Aparicio} \&
  {Hidalgo}}{2009}]{ah09}
{Aparicio} A.,  {Hidalgo} S.~L.,  2009, \mn@doi [\aj]
  {10.1088/0004-6256/138/2/558}, \href
  {http://adsabs.harvard.edu/abs/2009AJ....138..558A} {138, 558}

\bibitem[\protect\citeauthoryear{{Bekki} \& {Tsujimoto}}{{Bekki} \&
  {Tsujimoto}}{2012}]{bt12}
{Bekki} K.,  {Tsujimoto} T.,  2012, \mn@doi [\apj]
  {10.1088/0004-637X/761/2/180}, \href
  {http://adsabs.harvard.edu/abs/2012ApJ...761..180B} {761, 180}
  
\bibitem[\protect\citeauthoryear{{Calzetti} et~al.,}{{Calzetti}
  et~al.}{2015}]{calzettietal2015}
{Calzetti} D.,  et~al., 2015, \mn@doi [\aj] {10.1088/0004-6256/149/2/51}, \href
  {http://adsabs.harvard.edu/abs/2015AJ....149...51C} {149, 51}

\bibitem[\protect\citeauthoryear{{Catelan} \& {de Freitas Pacheco}}{{Catelan}
  \& {de Freitas Pacheco}}{1992}]{catelanetal92}
{Catelan} M.,  {de Freitas Pacheco} J.~A.,  1992, \aap, \href
  {http://adsabs.harvard.edu/abs/1992A%26A...258L...5C} {258, L5}

\bibitem[\protect\citeauthoryear{{Choudhury}, {Subramaniam}  \&
  {Cole}}{{Choudhury} et~al.}{2016}]{chetal16}
{Choudhury} S.,  {Subramaniam} A.,   {Cole} A.~A.,  2016, \mn@doi [\mnras]
  {10.1093/mnras/stv2414}, \href
  {http://adsabs.harvard.edu/abs/2016MNRAS.455.1855C} {455, 1855}

\bibitem[\protect\citeauthoryear{{Cignoni}, {Cole}, {Tosi}, {Gallagher},
  {Sabbi}, {Anderson}, {Grebel}  \& {Nota}}{{Cignoni} et~al.}{2013}]{cetal13b}
{Cignoni} M.,  {Cole} A.~A.,  {Tosi} M.,  {Gallagher} J.~S.,  {Sabbi} E.,
  {Anderson} J.,  {Grebel} E.~K.,   {Nota} A.,  2013, \mn@doi [\apj]
  {10.1088/0004-637X/775/2/83}, 775, 83

\bibitem[\protect\citeauthoryear{{Cohen}}{{Cohen}}{1982}]{c82}
{Cohen} J.~G.,  1982, \mn@doi [\apj] {10.1086/160061}, \href
  {http://adsabs.harvard.edu/abs/1982ApJ...258..143C} {258, 143}

\bibitem[\protect\citeauthoryear{{Da Costa} \& {Armandroff}}{{Da Costa} \&
  {Armandroff}}{1990}]{da90}
{Da Costa} G.~S.,  {Armandroff} T.~E.,  1990, \mn@doi [\aj] {10.1086/115500},
  100, 162

  
  \bibitem[\protect\citeauthoryear{{de Boer} et~al.,}{{de Boer}
  et~al.}{2012}]{deboeretal12}
{de Boer} T.~J.~L.,  et~al., 2012, \mn@doi [\aap]
  {10.1051/0004-6361/201118378}, \href
  {http://adsabs.harvard.edu/abs/2012A%26A...539A.103D} {539, A103}

\bibitem[\protect\citeauthoryear{{del Pino}, {Hidalgo}, {Aparicio}, {Gallart},
  {Carrera}, {Monelli}, {Buonanno}  \& {Marconi}}{{del Pino}
  et~al.}{2013}]{delpinoetal13}
{del Pino} A.,  {Hidalgo} S.~L.,  {Aparicio} A.,  {Gallart} C.,  {Carrera} R.,
  {Monelli} M.,  {Buonanno} R.,   {Marconi} G.,  2013, \mn@doi [\mnras]
  {10.1093/mnras/stt833}, \href
  {http://adsabs.harvard.edu/abs/2013MNRAS.433.1505D} {433, 1505}

\bibitem[\protect\citeauthoryear{{del Pino}, {Aparicio}  \& {Hidalgo}}{{del
  Pino} et~al.}{2015}]{delpinoetal15}
{del Pino} A.,  {Aparicio} A.,   {Hidalgo} S.~L.,  2015, \mn@doi [\mnras]
  {10.1093/mnras/stv2174}, \href
  {http://adsabs.harvard.edu/abs/2015MNRAS.454.3996D} {454, 3996}


  
  
\bibitem[\protect\citeauthoryear{{Geisler} \& {Sarajedini}}{{Geisler} \&
  {Sarajedini}}{1999}]{gs99}
{Geisler} D.,  {Sarajedini} A.,  1999, \mn@doi [\aj] {10.1086/300668}, 117, 308

\bibitem[\protect\citeauthoryear{{Geisler}, {Piatti}, {Bica}  \&
  {Clari{\'a}}}{{Geisler} et~al.}{2003}]{getal03}
{Geisler} D.,  {Piatti} A.~E.,  {Bica} E.,   {Clari{\'a}} J.~J.,  2003, \mn@doi
  [\mnras] {10.1046/j.1365-8711.2003.06408.x}, 341, 771

\bibitem[\protect\citeauthoryear{{Girardi} \& {Salaris}}{{Girardi} \&
  {Salaris}}{2001}]{gs01}
{Girardi} L.,  {Salaris} M.,  2001, \mn@doi [\mnras]
  {10.1046/j.1365-8711.2001.04084.x}, \href
  {http://adsabs.harvard.edu/abs/2001MNRAS.323..109G} {323, 109}

\bibitem[\protect\citeauthoryear{{Hendricks}, {Koch}, {Walker}, {Johnson},
  {Pe{\~n}arrubia}  \& {Gilmore}}{{Hendricks} et~al.}{2014}]{hendricksetal14}
{Hendricks} B.,  {Koch} A.,  {Walker} M.,  {Johnson} C.~I.,  {Pe{\~n}arrubia}
  J.,   {Gilmore} G.,  2014, \mn@doi [\aap] {10.1051/0004-6361/201424645},
  \href {http://adsabs.harvard.edu/abs/2014A%26A...572A..82H} {572, A82}

\bibitem[\protect\citeauthoryear{{Hidalgo} et~al.,}{{Hidalgo}
  et~al.}{2011}]{hidalgoetal11}
{Hidalgo} S.~L.,  et~al., 2011, \mn@doi [\apj] {10.1088/0004-637X/730/1/14},
  \href {http://adsabs.harvard.edu/abs/2011ApJ...730...14H} {730, 14}

\bibitem[\protect\citeauthoryear{{Kelson} et~al.,}{{Kelson}
  et~al.}{2014}]{kelsonetal2014}
{Kelson} D.~D.,  et~al., 2014, \mn@doi [\apj] {10.1088/0004-637X/783/2/110},
  \href {http://adsabs.harvard.edu/abs/2014ApJ...783..110K} {783, 110}

\bibitem[\protect\citeauthoryear{{Marigo}, {Girardi}, {Bressan}, {Groenewegen},
  {Silva}  \& {Granato}}{{Marigo} et~al.}{2008}]{metal08}
{Marigo} P.,  {Girardi} L.,  {Bressan} A.,  {Groenewegen} M.~A.~T.,  {Silva}
  L.,   {Granato} G.~L.,  2008, \mn@doi [\aap] {10.1051/0004-6361:20078467},
  482, 883

\bibitem[\protect\citeauthoryear{{Olsen} \& {Salyk}}{{Olsen} \&
  {Salyk}}{2002}]{os02}
{Olsen} K.~A.~G.,  {Salyk} C.,  2002, \mn@doi [\aj] {10.1086/342739}, 124, 2045

\bibitem[\protect\citeauthoryear{{Ordo{\~n}ez} \& {Sarajedini}}{{Ordo{\~n}ez}
  \& {Sarajedini}}{2015}]{os15}
{Ordo{\~n}ez} A.~J.,  {Sarajedini} A.,  2015, \mn@doi [\aj]
  {10.1088/0004-6256/149/6/201}, \href
  {http://adsabs.harvard.edu/abs/2015AJ....149..201O} {149, 201}

\bibitem[\protect\citeauthoryear{{Paczy{\'n}ski} \& {Stanek}}{{Paczy{\'n}ski}
  \& {Stanek}}{1998}]{ps98}
{Paczy{\'n}ski} B.,  {Stanek} K.~Z.,  1998, \mn@doi [\apjl] {10.1086/311181},
  \href {http://adsabs.harvard.edu/abs/1998ApJ...494L.219P} {494, L219}

\bibitem[\protect\citeauthoryear{{Phelps}, {Janes}  \& {Montgomery}}{{Phelps}
  et~al.}{1994}]{petal94}
{Phelps} R.~L.,  {Janes} K.~A.,   {Montgomery} K.~A.,  1994, \mn@doi [\aj]
  {10.1086/116920}, 107, 1079

\bibitem[\protect\citeauthoryear{{Piatti}}{{Piatti}}{2010}]{p10}
{Piatti} A.~E.,  2010, \mn@doi [\aap] {10.1051/0004-6361/201014216}, 513, L13

\bibitem[\protect\citeauthoryear{{Piatti}}{{Piatti}}{2011}]{p11a}
{Piatti} A.~E.,  2011, \mn@doi [\mnras] {10.1111/j.1745-3933.2011.01139.x},
  418, L40

\bibitem[\protect\citeauthoryear{{Piatti}}{{Piatti}}{2012}]{p12a}
{Piatti} A.~E.,  2012, \mn@doi [\mnras] {10.1111/j.1365-2966.2012.20684.x},
  422, 1109

\bibitem[\protect\citeauthoryear{{Piatti} \& {Geisler}}{{Piatti} \&
  {Geisler}}{2013}]{pg13}
{Piatti} A.~E.,  {Geisler} D.,  2013, \mn@doi [\aj]
  {10.1088/0004-6256/145/1/17}, 145, 17

\bibitem[\protect\citeauthoryear{{Piatti}, {Geisler}, {Bica}  \&
  {Clari{\'a}}}{{Piatti} et~al.}{2003a}]{petal03}
{Piatti} A.~E.,  {Geisler} D.,  {Bica} E.,   {Clari{\'a}} J.~J.,  2003a,
  \mn@doi [\mnras] {10.1046/j.1365-8711.2003.06727.x}, 343, 851

\bibitem[\protect\citeauthoryear{{Piatti}, {Bica}, {Geisler}  \&
  {Clari{\'a}}}{{Piatti} et~al.}{2003b}]{petal03b}
{Piatti} A.~E.,  {Bica} E.,  {Geisler} D.,   {Clari{\'a}} J.~J.,  2003b,
  \mn@doi [\mnras] {10.1046/j.1365-8711.2003.06887.x}, 344, 965

\bibitem[\protect\citeauthoryear{{Piatti}, {Sarajedini}, {Geisler}, {Clark}  \&
  {Seguel}}{{Piatti} et~al.}{2007}]{petal07d}
{Piatti} A.~E.,  {Sarajedini} A.,  {Geisler} D.,  {Clark} D.,   {Seguel} J.,
  2007, \mn@doi [\mnras] {10.1111/j.1365-2966.2007.11604.x}, 377, 300

\bibitem[\protect\citeauthoryear{{Piatti}, {Geisler}  \& {Mateluna}}{{Piatti}
  et~al.}{2012}]{pietal12}
{Piatti} A.~E.,  {Geisler} D.,   {Mateluna} R.,  2012, \mn@doi [\aj]
  {10.1088/0004-6256/144/4/100}, 144, 100

\bibitem[\protect\citeauthoryear{{Piatti}, {del Pino}, {Aparicio}  \&
  {Hidalgo}}{{Piatti} et~al.}{2014}]{petal14c}
{Piatti} A.~E.,  {del Pino} A.,  {Aparicio} A.,   {Hidalgo} S.~L.,  2014,
  \mn@doi [\mnras] {10.1093/mnras/stu1254}, 443, 1748

\bibitem[\protect\citeauthoryear{{Rey}, {Lee}, {Ree}, {Joo}, {Sohn}  \&
  {Walker}}{{Rey} et~al.}{2004}]{reyetal04}
{Rey} S.-C.,  {Lee} Y.-W.,  {Ree} C.~H.,  {Joo} J.-M.,  {Sohn} Y.-J.,
  {Walker} A.~R.,  2004, \mn@doi [\aj] {10.1086/380942}, \href
  {http://adsabs.harvard.edu/abs/2004AJ....127..958R} {127, 958}

\bibitem[\protect\citeauthoryear{{Rubele} et~al.,}{{Rubele}
  et~al.}{2012}]{retal12}
{Rubele} S.,  et~al., 2012, \mn@doi [\aap] {10.1051/0004-6361/201117863}, 537,
  A106

\bibitem[\protect\citeauthoryear{{Rubele} et~al.,}{{Rubele}
  et~al.}{2015}]{retal15}
{Rubele} S.,  et~al., 2015, \mn@doi [\mnras] {10.1093/mnras/stv141}, \href
  {http://adsabs.harvard.edu/abs/2015MNRAS.449..639R} {449, 639}

\bibitem[\protect\citeauthoryear{{Sabbi} et~al.,}{{Sabbi}
  et~al.}{2009}]{setal09}
{Sabbi} E.,  et~al., 2009, \mn@doi [\apj] {10.1088/0004-637X/703/1/721}, 703,
  721

\bibitem[\protect\citeauthoryear{{Savino}, {Salaris}  \& {Tolstoy}}{{Savino}
  et~al.}{2015}]{savinoetal15}
{Savino} A.,  {Salaris} M.,   {Tolstoy} E.,  2015, \mn@doi [\aap]
  {10.1051/0004-6361/201527072}, \href
  {http://adsabs.harvard.edu/abs/2015A%26A...583A.126S} {583, A126}

\bibitem[\protect\citeauthoryear{{Subramaniam}}{{Subramaniam}}{2003}]{s03}
{Subramaniam} A.,  2003, \mn@doi [\apjl] {10.1086/380556}, \href
  {http://adsabs.harvard.edu/abs/2003ApJ...598L..19S} {598, L19}

\bibitem[\protect\citeauthoryear{{Valenti}, {Ferraro}  \& {Origlia}}{{Valenti}
  et~al.}{2004}]{valentietal04}
{Valenti} E.,  {Ferraro} F.~R.,   {Origlia} L.,  2004, \mn@doi [\mnras]
  {10.1111/j.1365-2966.2004.07861.x}, \href
  {http://adsabs.harvard.edu/abs/2004MNRAS.351.1204V} {351, 1204}

\bibitem[\protect\citeauthoryear{{VandenBerg}, {Stetson}  \&
  {Brown}}{{VandenBerg} et~al.}{2015}]{vandenbergetal15}
{VandenBerg} D.~A.,  {Stetson} P.~B.,   {Brown} T.~M.,  2015, \mn@doi [\apj]
  {10.1088/0004-637X/805/2/103}, \href
  {http://adsabs.harvard.edu/abs/2015ApJ...805..103V} {805, 103}

\bibitem[\protect\citeauthoryear{{Wang}, {Strigari}, {Lovell}, {Frenk}  \&
  {Zentner}}{{Wang} et~al.}{2016}]{wangetal16}
{Wang} M.-Y.,  {Strigari} L.~E.,  {Lovell} M.~R.,  {Frenk} C.~S.,   {Zentner}
  A.~R.,  2016, \mn@doi [\mnras] {10.1093/mnras/stw220}, \href
  {http://adsabs.harvard.edu/abs/2016MNRAS.457.4248W} {457, 4248}


\makeatother
\end{thebibliography}

\input{paper.bbl}








\bsp	
\label{lastpage}
\end{document}